\documentclass[12pt]{article}
\usepackage{color}
\newcommand{\beq}{\begin{equation}}
\newcommand{\eeq}{\end{equation}}
\newcommand{\beqa}{\begin{eqnarray}}
\newcommand{\eeqa}{\end{eqnarray}}
\newcommand{\beqar}{\begin{eqnarray*}}
\newcommand{\eeqar}{\end{eqnarray*}}

\newcommand{\al}{\alpha}
\newcommand{\be}{\beta}

\def\spa          {\ \ \ }
\def\non          {\nonumber}
\def\ha           {\mbox{$\frac{1}{2}$}}

\def\s  {\sigma}
\def\spa          {\ \ \ }
\def\mand         {\spa\mbox{and}\spa}

\def\Tr           {\mbox{\rm Tr}\,}

\def\cd           {{\cdot}}
\def\ran          {\rangle}
\def\lan          {\langle}
\def\fsC    {C\!\!\!\!/\,}
\def\fsH    {H\!\!\!\!/\,}
\newcommand{\del}{\delta}

\newcommand{\eps}{\epsilon}
\newcommand{\ga}{\gamma}

\newcommand{\inn}{\!\cdot\!}

\newcommand{\lam}{\lambda}

\newcommand{\labell}[1]{\label{#1}} %{\label{#1}} %
\newcommand{\reef}[1]{(\ref{#1})}
\newcommand\prt{\partial}
\newcommand\veps{\varepsilon}

\newcommand\cL{{\cal L}}
\newcommand\cD{{\cal D}}

\newcommand\bz{\bar{z}}

\def\sst#1{{\scriptscriptstyle #1}}
\def\0{{\sst{(0)}}}
\def\1{{\sst{(1)}}}
\def\2{{\sst{(2)}}}
\def\3{{\sst{(3)}}}
\def\4{{\sst{(4)}}}
\def\5{{\sst{(5)}}}
\def\6{{\sst{(6)}}}
\def\7{{\sst{(7)}}}
\def\8{{\sst{(8)}}}

\begin{document}
\baselineskip 18pt%
\begin{titlepage}
\vspace*{1mm}%
\hfill
\vbox{

    \halign{#\hfil         \cr
   %       hep-th/yymmnnn\cr
        % ICTP-PH-TH/2012-xyz\cr
         %IPM/P-2010/003  \cr
         %  CPHT RR-xxx .yyzz \cr
           } % end of \halign
      }  % end of \vbox
\vspace*{8mm}
\vspace*{8mm}%

\center{ {\bf \Large  On New Bulk Singularity Structures, RR Couplings in 
\\
Asymmetric Picture and Their All Order $\alpha'$ Corrections}}\vspace*{3mm} \centerline{{\Large {\bf  }}}
\vspace*{5mm}
\begin{center}
{Ehsan Hatefi $^{a,b,1}$}

\vspace*{0.6cm}{\small $^{a}$  Centre for Research in String Theory, School of Physics and Astronomy,
\\
Queen Mary University of London, Mile End Road, London E1 4NS, United Kingdom},
\vskip.06in
%,\\and
{ $^{b}$ Institute for Theoretical Physics, TU Wien
\\
Wiedner Hauptstrasse 8-10/136, A-1040 Vienna, Austria }
\footnote{e.hatefi@qmul.ac.uk,ehsan.hatefi@tuwien.ac.at,ehsan.hatefi@cern.ch,ehsanhatefi@gmail.com}

\vspace*{.3cm}
\end{center}
\begin{center}{\bf Abstract}\end{center}
\begin{quote}

We have analyzed in detail  four and five point functions of the string theory amplitudes, including a closed string Ramond-Ramond (RR) in an asymmetric picture and either two or three  transverse scalar fields in both IIA and IIB. The complete forms of these S-matrices are derived and  these asymmetric S-matrices are also compared with their own symmetric results. This leads us to explore two different kinds of bulk singularity structures as well as various new couplings in asymmetric picture of the amplitude in type II string theory. All order $\alpha'$ higher derivative corrections to these new couplings have been discovered as well. Several remarks for  these two new bulk singularity structures and for contact interactions of the S-Matrix  have also been made.

 \end{quote}
\end{titlepage}%--------------------------------------------------------------------

 %%%%%%%%%%%%%%%%%%%%%%%%%%%%%%%%%%%%%%%%%%%%%%%%%%%%%%%%%%%%%%%%%%%%%%%%%%
\section{Introduction}

By now it is widely known that D-branes in super string theory play  the most important role in this area of research. Indeed there is no doubt they continue to have more contributions even to other topics of  high energy physics as well \cite{Polchinski:1995mt,Witten:1995im,Polchinski:1996na}\footnote{ Both super symmetric and non-super symmetric branes must be regarded as $p+1$ world volume dimensions in a flat empty space background, for which two different kinds of boundary conditions should have been taken into account \cite{Polchinski:1994fq}.}.

\vskip 0.1in

Although we have referred in \cite{Hatefi:2015ora} to various fascinating  papers  about the subject of string theory's effective actions, for the entire self-completeness, we point out some of the remarks that are of high importance to  the author as follows.

\vskip 0.1in

To our knowledge Myers in \cite{Myers:1999ps} discovered more or less the complete form of a single bosonic action that can be generalized for diverse D$_p$-brane systems. Not really consequently but after while we started finding out the generalization of that action with emphasis on exploring all order $\alpha'$ corrections to D-brane effective actions. This involves to deal with both Chern-Simons, DBI effective actions and mixed open-RR S-matrices. Besides those things , some of the new couplings and or Myers terms have been derived in \cite{Hatefi:2012zh}. It is also worth highlighting  the fact that applications to some of the new Myers terms have already been released in the literature.

\vskip 0.1in

For instance we have shown in \cite{Hatefi:2012sy}  that the so called $N^3$  entropy behaviour of near extremal M5 branes is reconstructed from superYang-Mills part with the particular role of Myers terms. In \cite{Hatefi:2012sy} it was seen that the presence of closed string interaction terms (Myers terms) is necessary to  reproduce that $N^3$ entropy growth, basically we had regenerated the leading $N^3$  entropy
behaviour from D0 quantum mechanics with Myers terms. The  leading growth just came
from the classical contribution. In this paper we also revealed that Myers terms come from just the closed string coupling to some lower dimensional branes
which lie inside of the branes that one started with. These lower dimensional
branes could be thought of just soliton solutions of the branes that one dealt with. Hence these Myers terms that can be found by world sheet computations are important  to actually explore $N^3$ entropy growth production and potentially might play role in M-theory  \cite{Horava:1995qa} as well. 

\vskip.1in

We have also introduced a new sort of Kaluza-Klein reduction method based on ADM decomposition. In particular in \cite{Hatefi:2012bp} we discussed  how the world volume theory appears from the supergravity side, where we applied the scheme to IIB supergravity that is reduced on a 5D hyperboloidal space, and clarified how one can gain either  AdS or dS brane world solutions by making further reduction to 4D.

\vskip 0.2in

Various remarks for D-brane anti D-brane system \cite{Michel:2014lva,Garousi:2007fk} involving their corrections \cite{Hatefi:2012cp} have been given. In addition to the efforts in \cite{Howe:2006rv}, the entire form of super symmetric Myers action has not been concluded yet. On the other hand, the action for a single brane  was understood in  \cite{Leigh:1989jq} where its generalization could be found in  \cite{Cederwall:1996pv}.

\vskip.1in

We invite the reader for the complete review of Chern-Simons effective actions and their remarks to have a look at \cite{li1996a,Douglas:1995bn,Green:1996dd}. One can find out a very brief review of all DBI and new Wess-Zumino terms of BPS branes in \cite{Hatefi:2010ik}.

 To observe all the three ways of exploring couplings in Effective Field Theory (EFT), (which are either Taylor or pull-back, Myers terms) for both super symmetric and non super symmetric cases we suggest  \cite{Hatefi:2012wj}.

\vskip 0.1in

Behind AdS/CFT there is a close relation  between an open and a closed string so one might be interested in gaining the mixed open-closed string amplitudes, which of particular interest to them in string theory is indeed the mixed RR potential (C-field)-open strings. In fact one may hope to address various issues (involving the AdS/CFT) by dealing with these mixed higher point functions of RR-scalar fields of type II. Let us just consider some of the works that are in correspondence either with S-matrix formalism in the presence of  D$_p$-branes or are related to D-brane physics applications  \cite{Hashimoto:1996bf}.

\vskip 0.2in

The paper is organized as follows. In section 2 , we talk about the conventions and then try to provide the complete calculations for  the four and five point functions of type IIA,IIB of   string theory including an RR in asymmetric picture and either two or three real transverse scalar fields. Indeed for $<V_{C^{-2}} V_{\phi^0} V_{\phi^0} >$ S-Matrix, we  modify the so called Wick-like rule to have the gauge invariance, particularly we show that in asymmetric picture of RR one finds out new term in the amplitude and hence new couplings in an EFT can be constructed out. In order to be able to produce all string contact interactions in an EFT, one needs to employ mixing couplings where the first scalar comes from Taylor expansion and the second one comes from pull-back method. We then find out  all order $\alpha'$ higher derivative correction to those couplings as well.

\vskip 0.2in

In the next sections we perform the entire analysis of a five point function of a C-field with three transverse scalar fields in zero picture, that is , we deal with $<V_{C^{-2}} V_{\phi^0} V_{\phi^0} V_{\phi^0}>$. We also compare the exact form of the S-matrix in asymmetric picture with its own result in symmetric picture $<V_{C^{-1}} V_{\phi^{-1}} V_{\phi^0} V_{\phi^0}>$ at both level of contact interaction and singularity structures. We  obtain various new contact interactions as well as two kinds of new bulk singularity structures with various new couplings in $<V_{C^{-2}} V_{\phi^0} V_{\phi^0} V_{\phi^0}>$ which are not appeared in $<V_{C^{-1}} V_{\phi^{-1}} V_{\phi^0} V_{\phi^0}>$.

 \vskip.1in

These two different kinds of  bulk singularity structures of string amplitude are $t,s,u$ as well as $(t+s+u)$-channels bulk singularity structures that can just be explored in an asymmetric picture of the amplitude. These bulk singularity structures carry momentum of RR in transverse directions  ( that is, $p^i, p^j, p^k$ terms). Note that these terms could have been derived if winding modes ($w^i, w^j, w^k$ terms) were allowed in the vertex operator of RR. However, in the vertex of RR  in ten dimensions of spacetime there are no winding modes in both symmetric and asymmetric picture of RR. 
Indeed these terms of the amplitude whose momenta of RR are carried in transverse direction cannot be obtained even by T-duality
transformation in flat ten dimensions of space-time. Hence the  presence of RR makes computations complicated as was explained in \cite{Hatefi:2012ve,Park:2008sg}.
Hence we just explore these new bulk singularities as well as new couplings in asymmetric picture of the amplitude. Indeed we are also able to produce these two different  bulk singularity structures of string amplitude in field theory  by taking into account various new couplings in effective field theory side as well.  
\vskip.1in

We also generalize all order $\alpha'$ higher derivative corrections to those new couplings that are appeared in an asymmetric picture of the amplitude.
 \vskip.1in

The important point must be noted. These new couplings are discovered by just scattering amplitude formalism in an antisymmetric picture and not any other tools such as T-duality can be employed to get to these couplings. Because these new couplings carry momentum of RR in transverse (or bulk) directions, while winding modes are not embedded in ten dimension of RR vertex operator. 
We then construct  all order bulk singularity structures of  $t,s,u, (t+s+u)$ channels in an EFT where  the universal conjecture of corrections \cite{Hatefi:2012rx} plays the fundamental ingredient in producing all the infinite singularity structures.

 \vskip.1in

 These new bulk singularity structures of string theory amplitude that are just shown up in $<V_{C^{-2}} V_{\phi^0} V_{\phi^0} V_{\phi^0}>$ can be generated by taking into account various new EFT couplings. All those new terms do carry the scalar products of momentum of RR in the bulk  with the polarization of scalar fields accordingly. We think that the importance of these results will be provided in future research topics , such as all order Myers effect and various other subjects in type II string theory \cite{Hatefi:2015mio}. We have also observed that at the level of EFT
 the  super gravity background fields in DBI action must be some functions of super Yang-Mills. Some particular Taylor expansion for the background fields should also be taken into account as was notified in Dielectric effect
 \cite{Myers:1999ps}.

\vskip 0.1in

\section{The $\phi^{0}-\phi^{0}-C^{-2}$ with all order $\alpha'$ corrections}

In this section we take into account some Conformal Field Theory (CFT) tools to get to the entire S-matrix $<V_{C^{-2}} V_{\phi^0}  V_{\phi^0}>$. Having expanded the elements of scattering amplitude and taken  some patters for string corrections, we would start generating $\alpha'$ corrections. Our S-matrix computations are valid at world-sheet level of four and five point functions at the disk level which covers both transverse and world volume directions.
Although it is impossible to address all the attempts that have been carried out in this area , we can highlight several efforts that are worth
considering  for super symmetric and non super symmetric cases \cite{Kennedy:1999nn,Barreiro:2013dpa,Liu:2001qa,Hatefi:2013mwa}.

\vskip.1in

One first needs to apply the general structure of vertices where in this paper we just insist on employing RR potential (which is a $C_{p+1}$-form field in asymmetric picture), therefore all the other two or three transverse scalars have to be considered at zero picture. The vertex of RR in asymmetric picture was first proposed by \cite{Bianchi:1991eu}. A new paper about picture changing operators has been recently released \cite{Sen:2015hia}, however, to our knowledge it is not 
understood how to deal with  all RR closed-open string amplitudes.  That is why we try to come up with direct calculations, although the computations in an asymmetric picture is very long and tedious. It would be very nice to work out more to actually understand whether or not the proposal in \cite{Sen:2015hia} can be applied to higher point functions of string theory including RR (in an asymmetric picture) and scalar fields.
The three point function of a closed string RR and a transverse scalar field (describing oscillation of brane) in both symmetric and asymmetric picture of RR has been accordingly computed in detail \cite{Hatefi:2015gwa}. Let us as a warm-up just mention the results
\beqa
{\cal A}^{C^{-1}\phi^{-1}} &=& 2^{-1/2}\Tr
(P_{-}\fsH_{(n)}M_p\gamma^{i})\xi_{1i} \labell{ee}\ .
\eeqa
also
\beqa
{\cal A}^{\phi^{0},C^{-2}} &=& \bigg[-i p^i\Tr(P_{-}\fsC_{(n-1)}M_p)+ik_{1a}\Tr(P_{-}\fsC_{(n-1)}M_p\Gamma^{ia})\bigg]\xi_{1i}\label{ham12}
\eeqa
where definitions could be read off from \cite{Hatefi:2015gwa}. We have chosen the following notations for  entire ten dimensional space-time, world volume and transverse directions appropriately 
\beqa
\mu,\nu &=& 0, 1,..., 9 \nonumber\\
 a, b, c &=& 0, 1,..., p\nonumber\\
i,j &=& p + 1,...,9 \nonumber\eeqa

To be able to produce ${\cal A}^{\phi^{0},C^{-2}}$ in an EFT part,  we first apply to the second term of \reef{ham12} momentum conservation on world volume  $(k_1^{a} + p^{a} =0)$ and in particular consider the following Bianchi identity

   \beqa
   p^a\eps^{a_{0}\cdots a_{p-1}a}
%C^i_{a_{0}\cdots a_{p-1}}
 =0
   \label{ham22}\eeqa

so that the 2nd term in  \reef{ham12} has zero contribution to the S-Matrix. The 1st term in \reef{ham12} can be produced in an EFT by considering    Taylor expansion of a real scalar field through RR coupling as follows
\beqa
S_1
&=& \frac{(2\pi\alpha')\mu_p}{(p+1)!}\int d^{p+1}\sigma (\veps^v)^{a_0\cdots a_{p}}
\Tr(\phi^i)
\partial_i C_{a_0\cdots a_{p}}
\nonumber\eeqa
  \section{$<V_{C^{-2}} V_{\phi^0}  V_{\phi^0}>$ amplitude}
The four point function in an asymmetric picture of a RR and two real scalar fields  in zero picture can be done by
\beqa
{\cal A}^ {\phi^{0} \phi^{0}C^{-2}} & \sim & \int dx_1 dx_2 d^2z
 \lan V_\phi^{(0)}(x_1)V_\phi^{(0)}(x_2)
V_{RR}^{(-2)}(z,\bar{z})\ran\labell{cor10ti}\eeqa
The scalar field and RR vertex operators in symmetric and asymmetric  pictures are  given by
\beqa
V_{\phi}^{(0)}(x) &=& \xi_{i}\bigg(\partial
X^i(x)+\alpha' ik\cd\psi\psi^i(x)\bigg)e^{\alpha' ik\cd X(x)},
\nonumber\\
V_{\phi}^{(-2)}(x_1)&=& e^{-2\phi(x_1)}V_{\phi}^{(0)}(x_1) \nonumber\\
V_\phi^{(-1)}(x)&=&e^{-\phi(x)}\xi_i\psi^i(x)e^{2iq\inn X(x)} \nonumber\\
V_{RR}^{(-1)}(z,\bar{z})&=&(P_{-}\fsH_{(n)}M_p)^{\al\be}e^{-\phi(z)/2} S_{\al}(z)e^{i\frac{\alpha'}{2}p\cd X(z)}e^{-\phi(\bar{z})/2} S_{\be}(\bar{z}) e^{i\frac{\alpha'}{2}p\cd D \cd X(\bar{z})}\nonumber\\
V_{RR}^{(-2)}(z,\bar{z})&=&(P_{-}\fsC_{(n-1)}M_p)^{\al\be}e^{-3\phi(z)/2} S_{\al}(z)e^{i\frac{\alpha'}{2}p\cd X(z)}e^{-\phi(\bar{z})/2} S_{\be}(\bar{z}) e^{i\frac{\alpha'}{2}p\cd D \cd X(\bar{z})}\nonumber
\eeqa
where $k^2=p^2=0, k.\xi_1=0$. We also consider $x_{4}\equiv\ z=x+iy, x_{5}\equiv\bz=x-iy$ and  the definitions of  RR's field strength and projection operator  are
\begin{displaymath}
P_{-} =\ha (1-\ga^{11}), \quad
\fsH_{(n)} = \frac{a
_n}{n!}H_{\mu_{1}\ldots\mu_{n}}\ga^{\mu_{1}}\ldots
\ga^{\mu_{n}},
\non\end{displaymath}
For type IIA (IIB) $n=2,4$,$a_n=i$  ($n=1,3,5$,$a_n=1$) . Note that spinor notation reads off  
\beqa
(P_{-}\fsH_{(n)})^{\al\be} =
C^{\al\del}(P_{-}\fsH_{(n)})_{\del}{}^{\be}.
\nonumber
\eeqa
To make use of the holomorphic propagators  we apply the following doubling trick . Hence one uses  the following change of variables as

\begin{displaymath}
\tilde{X}^{\mu}(\bar{z}) \rightarrow D^{\mu}_{\nu}X^{\nu}(\bar{z}) \ ,
\spa
\tilde{\psi}^{\mu}(\bar{z}) \rightarrow
D^{\mu}_{\nu}\psi^{\nu}(\bar{z}) \ ,
\spa
\tilde{\phi}(\bar{z}) \rightarrow \phi(\bar{z})\,, \mand
\tilde{S}_{\al}(\bar{z}) \rightarrow M_{\al}{}^{\be}{S}_{\be}(\bar{z})
 \ ,
\non\end{displaymath}

with 
\begin{displaymath}
D = \left( \begin{array}{cc}
-1_{9-p} & 0 \\
0 & 1_{p+1}
\end{array}
\right) \ ,\,\, \mand
M_p = \left\{\begin{array}{cc}\frac{\pm i}{(p+1)!}\ga^{i_{1}}\ga^{i_{2}}\ldots \ga^{i_{p+1}}
\eps_{i_{1}\ldots i_{p+1}}\,\,\,\,{\rm for\, p \,even}\\ \frac{\pm 1}{(p+1)!}\ga^{i_{1}}\ga^{i_{2}}\ldots \ga^{i_{p+1}}\ga_{11}
\eps_{i_{1}\ldots i_{p+1}} \,\,\,\,{\rm for\, p \,odd}\end{array}\right.
\non\end{displaymath}
\vskip .1in
Finally we can use the following correlation functions for   $X^{\mu},\psi^\mu, \phi$,  fields as 
\begin{eqnarray}
\lan X^{\mu}(z)X^{\nu}(w)\ran & = & -\frac{\alpha'}{2}\eta^{\mu\nu}\log(z-w) \ , \non \\
\lan \psi^{\mu}(z)\psi^{\nu}(w) \ran & = & -\frac{\alpha'}{2}\eta^{\mu\nu}(z-w)^{-1} \ ,\non \\
%\lan c(z)c(w) \ran & = & (z-w) \ , \non \\
\lan\phi(z)\phi(w)\ran & = & -\log(z-w) \ .
\labell{prop2}\end{eqnarray}

Having set the Wick theorem, the  amplitude would be written down as
\beqa
&&\int dx_1 dx_2 dx_4 dx_5  (P_{-}\fsC_{(n-1)}M_p)^{\alpha\beta}  (x_{45})^{-3/4} \xi_{1i}\xi_{2j}  \bigg(K_1+K_2+K_3+K_4\bigg) K_5
 \nonumber\eeqa
 The exponential factors are
\beqa
K_5&=&|x_{12}|^{\alpha'^2 k_1.k_2}|x_{14}x_{15}|^{\frac{\alpha'^2}{2}k_1.p} |x_{24}x_{25}|^{\frac{\alpha'^2}{2}k_2.p}|x_{45}|^{\frac{\alpha'^2}{4}p.D.p},\nonumber\eeqa
In order to be able to explore all the above the fermionic correlators including the correlation functions  of various currents with spin operators  , one has to consider the Wick-Like rule \cite{Liu:2001qa}
and modify it as follows
\beqa
 \lan\psi^{\mu_1}
(x_1)...
\psi^{\mu_n}(x_n)S_{\al}(x_4)S_{\be}(x_5)\ran&\!\!\!\!=\!\!\!\!
&\frac{1}{2^{n/2}}
\frac{(x_4-x_5)^{n/2-5/4}}
{|x_1-x_4|...|x_n-x_4|}\left[(\Gamma^{\mu_n...\mu_1}
C^{-1})_{\al\be}\right.\nonumber\\&&+
\lan\psi^{\mu_1}(x_1)\psi^{\mu_2}(x_2)\ran(\Gamma^{\mu_n...\mu_3}
C^{-1})_{\al\be}
\pm perms\nonumber\\&&+\lan\psi^{\mu_1}(x_1)\psi^{\mu_2}(x_2)\ran
\lan\psi^{\mu_3}(x_3)\psi^{\mu_4}(x_4)\ran(\Gamma^{\mu_n...\mu_5}
C^{-1})_{\al\be}\nonumber\\&&\left.
\pm perms+\cdots\right]\labell{wicklike1}\eeqa

 where  one has to consider all various contractions to make sense of the corrected S-matrix elements. The  important point one has to hold, is that in all the above equations, $\Gamma^{\mu_{n}...\mu_{1}}$ must be  antisymmetric with respect to all the 
 gamma matrices. The $x_i's$  are real and other important point is as follows. Apart from considering co-cycles in the fermionic two point functions, the Wick-like formula must be modified with a minus sign, in order to respect the gauge invariance of the higher point function of BPS or non-BPS S-matrices. Hence, the corrected Wick-like formula for the definition of two point function is
  \beqar\lan\psi^{\mu}(x_{1})\psi^{\nu}(x_{2})\ran &=&-\alpha'\eta^{\mu\nu}{\frac {Re[x_{14} x_{25}]}{x_{12} x_{45}}} =-\alpha'\eta^{\mu\nu}{\frac {Re[(x_{1}-x_4)(x_{2}-x_5)]}{(x_{1}-x_{2})(x_4-x_5)}}\labell{wicklike12}\eeqar

To elaborate and explain how exactly these signs guarantee gauge invariance, let us not consider the minus sign in  Wick-Like formula (and dropped out co-cycles in the fermionic functions) and focus on the amplitude of an RR, a gauge field and two real tachyons. Having set that, the following
correlation function  
$<:S_{\al}(x_4):S_{\be}(x_5)
:{\psi^a(x_1)}:{\psi^b(x_2)}:{\psi^c(x_3)}:>$ would be found 
\beqa
 2^{-3/2}(x_{14}x_{15}x_{24}x_{25} x_{34}x_{35})^{-1/2}  (x_{45})^{1/4}\bigg[ ( \Gamma^{cba}C^{-1})_{\alpha\beta}+2\eta^{ab}(\gamma^{c}C^{-1})_{\alpha\beta}\frac{Re[x_{14}x_{25}]}{x_{12}x_{45}}
\nonumber\\+2\eta^{ac}(\gamma^{b}C^{-1})_{\alpha\beta}\frac{Re[x_{14}x_{35}]}{x_{13}x_{45}}
+2\eta^{bc}(\gamma^{a
}C^{-1})_{\alpha\beta} \frac{Re[x_{24}x_{35}]}{x_{23}x_{45}}
\bigg]\labell{bby}\eeqa
where Mandelstam variables are 
$s=-(k_1+k_3)^2, t=-(k_1+k_2)^2, u=-(k_2+k_3)^2$.
 Given \reef{bby}, the final form of the S-matrix is 
\beqa
{\cal A}^{C^{-1} A^{-1}T^0T^0}&=&\frac{i\mu_p}{2\sqrt{2\pi}}\left[\Tr\bigg((P_{-}\fsH_{(n)}M_p)
(k_3.\ga)(k_2.\ga)(\xi.\ga)\bigg)I\delta_{p,n+2}
+\Tr\bigg((P_{-}\fsH_{(n)}M_p)
\ga^{a}\bigg)J\delta_{p,n}\right. \nonumber\\&&\left.\times
\bigg\{-k_{2a}(t+1/4)(2\xi.k_{3})
+k_{3a}(s+1/4)(2\xi.k_{2})-\xi_a(s+1/4)(t+1/4)\bigg\}\right]\labell{gen}\eeqa
Now clearly if we replace $\xi_{1a}\rightarrow k_{1a}$ in \reef{bby} then one gets to know that the amplitude does not vanish any more. Because after replacing  $\xi_{1a}\rightarrow k_{1a}$ the amplitude is proportional to
$(s+1/4)(t+1/4) (-k_{1a}-k_{3a}+k_{2a})$ which is not zero and hence it is not gauge invariant any more. So one needs to consider the minus sign in co-cycles as well as the minus sign in Wick-Like formula\footnote{where $I,\,J$ are  also given in  \cite{Garousi:2007fk}}.

Now we apply the generalisation of Wick-like rule to amplitude so that one is able to derive  all the correlators as
\beqa
 K_1&=& \bigg(-\eta^{ij} x_{12}^{-2}+(ip^{i}\frac{x_{54}}{x_{14}x_{15}})(ip^{j}\frac{x_{54}}{x_{24}x_{25}}) \bigg)   (x_{45})^{-5/4}( C^{-1})_{\alpha\beta}\nonumber\\&&
K_2 =(ip^{i}\frac{x_{54}}{x_{14}x_{15}}) ik_{2b} (x_{24}x_{25})^{-1} (x_{45})^{-1/4} ( \Gamma^{jb}C^{-1})_{\alpha\beta}\nonumber\\&&
K_3= (ip^{j}\frac{x_{54}}{x_{24}x_{25}}) ik_{1a} (x_{14}x_{15})^{-1} (x_{45})^{-1/4} ( \Gamma^{ia}C^{-1})_{\alpha\beta}\nonumber\\&&
K_4=-k_{1a}k_{2b} (x_{14}x_{15}x_{24}x_{25})^{-1}  (x_{45})^{3/4} \bigg[ ( \Gamma^{jbia}C^{-1})_{\alpha\beta}+\frac{Re[x_{14}x_{25}]}{x_{12}x_{45}}
\nonumber\\&&\times
(2\eta^{ab}(\Gamma^{ji}C^{-1})_{\alpha\beta}+2\eta^{ij}(\Gamma^{ba
}C^{-1})_{\alpha\beta}) -4\eta^{ab}\eta^{ij}(C^{-1})_{\alpha\beta}(\frac{Re[x_{14}x_{25}]}{x_{12}x_{45}})^2\bigg] \label{mion}\eeqa
By applying \reef{mion}  into this four point amplitude we can easily determine that the  S-matrix is $SL(2,R)$ invariant.
 We do the proper gauge fixing  as  $(x_1,x_2,z,\bar z)=(x,-x,i,-i)$, taking $t=- \frac{\alpha'}{2} (k_1+k_2)^2$ to get to the S-matrix as
 \beqa
 {\cal A}^{\phi^{0}\phi^{0}C^{-2}}&=&-(2i)^{-2t}\xi_{1i}\xi_{2j} (P_{-}\fsC_{(n-1)}M_p)^{\alpha\beta}\int_{-\infty}^{\infty}dx (1+x^2)^{2t-1} (2x)^{-2t}
\nonumber\\&&\times \bigg\{(2i)^{-1}(C^{-1})_{\alpha\beta}
 \bigg(-\eta^{ij} (1+x^2)^{2} (2x)^{-2}+4 p^i p^j\bigg)\nonumber\\&&
+2i \bigg(p^i k_{2b} (\Gamma^{jb}C^{-1})_{\alpha\beta}+ p^j k_{1a} (\Gamma^{ia}C^{-1})_{\alpha\beta}\bigg)
\nonumber\\&&
-k_{1a}k_{2b} (2i) \bigg[(\Gamma^{jbia}C^{-1})_{\alpha\beta}+\bigg(2\eta^{ab}(\Gamma^{ji}C^{-1})_{\alpha\beta}+2\eta^{ij}(\Gamma^{ba
}C^{-1})_{\alpha\beta}\bigg) \frac{1-x^2}{4ix}
\nonumber\\&&
-4\eta^{ab}\eta^{ij}(C^{-1})_{\alpha\beta}(\frac{1-x^2}{4ix})^2\bigg]\bigg\}\label{lop}\eeqa
where the sixth and seventh terms  do not have any contribution to the S-matrix because the integration is taken on the whole space meanwhile the integrand is odd so the result  vanishes. More crucially, after using kinematical relations  we come to know that the sum of the first term  and the last term of \reef{lop} is zero. 
%If we use momentum conservation along the world volume of brane $(k_1+k_2+p)^a=0$ then one finds out the summation of  the 3rd and 4th term of \reef{lop} as follows

%\beqa
% {\cal A}^{\phi^{0}\phi^{0}C^{-2}}_1&=& \bigg[-p^i k_{1a}\Tr(P_{-}\fsC_{(n-1)}M_p \Gamma^{ja}) +p^j k_{1a}\Tr(P_{-}\fsC_{(n-1)}M_p \Gamma^{ia})-p^b p^i\Tr(P_{-}\fsC_{(n-1)}M_p \Gamma^{jb})\bigg] \nonumber\\&&
% \times(2i)^{-2t+1}\xi_{1i}\xi_{2j} \sqrt(\pi)  \frac{ \Gamma(-t+\frac{1}{2})}{ \Gamma(-t+1)}\label{lop738}\eeqa

%\beqa
%-p^i \epsilon^{a_{0}...a_{p-1}a}H^{j}_{a_{0}...a_{p-1}}+p^j \epsilon^{a_{0}...a_{p-1}a}H^{i}_{a_{0}...a_{p-1}}=0\label{esi}\eeqa
 
One now explores  the  final answer for  the 3rd and 4th term of \reef{lop}  as follows

\beqa
 {\cal A}^{\phi^{0}\phi^{0}C^{-2}}_1&=& (2i)^{-2t+1}\xi_{1i}\xi_{2j} \sqrt(\pi)  \frac{ \Gamma(-t+\frac{1}{2})}{ \Gamma(-t+1)}
\nonumber\\&&\times\bigg[p^i k_{2b}\Tr(P_{-}\fsC_{(n-1)}M_p \Gamma^{jb}) +p^j k_{1a}\Tr(P_{-}\fsC_{(n-1)}M_p \Gamma^{ia})\bigg]
 \label{lop7399}\eeqa

%\beqa
 %{\cal A}^{\phi^{0}\phi^{0}C^{-2}}_1&=& (2i)^{-2t+1}\xi_{1i}\xi_{2j} \sqrt(\pi) k_{1a} \frac{ \Gamma(-t+\frac{1}{2})}{ \Gamma(-t+1)}
 %\nonumber\times\\&&\bigg[-p^i \Tr(P_{-}\fsC_{(n-1)}M_p \Gamma^{jb}) +p^j \Tr(P_{-}\fsC_{(n-1)}M_p \Gamma^{ia})\bigg]
% \label{lop738}\eeqa

Eventually one can find out the result for the 2nd term of asymmetric S-matrix as follows

 \beqa
 {\cal A}^{\phi^{0}\phi^{0}C^{-2}}_2&=& -(2i)^{-2t}\xi_{1i}\xi_{2j} \sqrt(\pi)\Tr(P_{-}\fsC_{(n-1)}M_p) \bigg\{
 %(\frac{i}{2}-\frac{i}{2}) \eta^{ij}\frac{\sqrt(\pi) \Gamma(-t-\frac{1}{2})}{ \Gamma(-t)}\bigg\}
 -2i p^i p^j\frac{ \Gamma(-t+\frac{1}{2})}{ \Gamma(-t+1)} \bigg\}\label{lop78}\eeqa

The trace is non-zero for $p+1=n-1$ case and \reef{lop78} does not include any poles, because the expansion is low energy expansion ($t\rightarrow 0$). It has been emphasized in \cite{Hatefi:2012wj} that for a particular string S-matrix involving scalar fields one needs to take into account three different ways to actually re-generate all the string couplings in an EFT side. The first way was imposed by Myers in \cite{Myers:1999ps}, while the second and the third ways of EFT have been completely mentioned in \cite{Hatefi:2012wj} to be  either pull-back method or Taylor expansion  of the scalar fields.
\vskip.1in

This S-matrix in symmetric picture can be readily computed as
\beqa {\cal A}^{C^{-1}\phi^{-1}\phi^{0}}&=&i(2i)^{-2t+1}\int_{-\infty}^{\infty} dx (x^2+1)^{2t-1} (2x)^{-2t} (\xi_{1i} \xi_{2j} 2^{-1/2})\nonumber\\&&\times
\bigg[p^j\Tr(P_{-}\fsH_{(n)}M_p\gamma^{i})
-k_{2a}\Tr(P_{-}\fsH_{(n)}M_p\Gamma^{jai})\bigg]
\labell{mm515}
\eeqa
%%%
 %where$t=\frac{-\alpha'}{2}(k_1+k_2)^2$.

Carrying out the integrals explicitly and using
momentum conservation, we obtain
%%%
\beqa {\cal A}^{C^{-1}\phi^{-1}\phi^{0}}&=& (2\xi_{1i} \xi_{2j} 2^{-1/2}\pi^{1/2})\frac{\Gamma(-t+\frac{1}{2})}{\Gamma(-t+1)}\nonumber\\&&\times
\bigg[p^j\Tr(P_{-}\fsH_{(n)}M_p\gamma^{i})
-k_{2a}\Tr(P_{-}\fsH_{(n)}M_p\Gamma^{jai})\bigg]
\labell{mm567}
\eeqa
The closed form of the  expansion is
\beqa
\sqrt{\pi}\frac{\Gamma(-t+\frac{1}{2})}{\Gamma(-t+1)}&=&\pi \sum_{n=-1}^{\infty} c_n(t)^{n+1}\label{infinite}
\eeqa
where  all $c_n$ are related to higher derivative corrections of the scalar fields through either Taylor expansions or Pull-back of branes.
\beqa
c_{-1}=1,\quad c_{0}=2ln(2),\quad c_{1}=\frac{\pi^2}{6}+2ln(2)^2
\eeqa
Now we work out the related super Yang-Mills  vertices to reconstruct all string couplings to all orders in  $\alpha'$.
In order to produce all infinite contact interactions for first term of \reef{mm567} and also to produce ${\cal A}^{\phi^{0}\phi^{0}C^{-2}}_2$ part, one needs to  deal with the Taylor expansion of the two real scalar fields through an
   RR ($p$+1)-form field as
 \beqa S_2
&=&i\frac{(2\pi\alpha')^2\mu_p}{2(p+1)!}\int d^{p+1}\sigma
(\veps^v)^{a_0\cdots a_{p}} \Tr(\Phi^i\Phi^j)
\prt_{i}\prt_{j}C^{(p+1)}_{a_0\cdots a_{p}} \nonumber\eeqa
and subsequently all order $\alpha'$ corrections can be derived as
 %%%%
%\beqa S^{(3)} &=&i\frac{\lambda^2\mu_p}{2(p+1)!}\int d^{p+1}\sigma
%(\veps^v)^{a_0\cdots a_{p}} \Tr(\Phi^i\Phi^j)
%\prt_{j}H^{(p+2)}_{ia_0\cdots a_{p}}(\sigma) \label{S21} \eeqa
%%%

\beqa S_3  &=&\frac{(2\pi\alpha')^2\mu_p}{2(p+1)!}\int d^{p+1}\sigma
(\veps^v)^{a_0\cdots a_{p}}p^j H^{(p+2)}_{ia_0\cdots
a_{p}}\nonumber\\&&\times
\sum_{n=-1}^{\infty}c_{n}(\alpha')^{n+1}\Tr(\prt_{a_{1}}...\prt_{a_{n+1}}\Phi^i\prt^{a_{1}}...\prt^{a_{n+1}}\Phi^j)
\label{S122} \eeqa

In the meantime to be able to produce at the leading order the second term of \reef{mm567} as well as the 5th term of  ${\cal A}^{\phi^{0}\phi^{0}C^{-2}}$, one must employ the so called Pull-back formalism as follows
\beqa S_4  &=&i \frac{(2\pi\alpha')^2\mu_p} {2(p-1)!}\int d^{p+1}\sigma
(\veps^v)^{a_0\cdots a_{p}} \Tr(D_{a_{0}}\Phi^iD_{a_{1}}\Phi^j)
C^{(p+1)}_{ija_2\cdots a_{p}} \label{pull} \eeqa
Essentially one can insist on producing all infinite corrections by imposing  the higher derivative corrections to the pull-back  and fix all their coefficients without any ambiguities in string theory  as below
 \beqa S_5  &=&i \frac{(2\pi\alpha')^2\mu_p} {2(p)!}\int
d^{p+1}\sigma (\veps^v)^{a_0\cdots a_{p}}
\sum_{n=-1}^{\infty}c_{n}(\alpha')^{n+1}\Tr(\prt_{a_{1}}...\prt_{a_{n+1}}
D_{a_{0}}\Phi^i \prt^{a_{1}}...\prt^{a_{n+1}} \Phi^j)
 H^{(p+2)}_{ija_1\cdots a_{p}}
\nonumber
\eeqa
%%%
There is no external gauge field in our S-matrix so one could propose the covariant derivatives of scalar fields to above higher derivative couplings to be able to keep track of the gauge invariance  of the S-matrix as well.

\subsection{ The other RR couplings of type II string theory}

Now in order to be able to produce  the term  that has been explicitly appeared by asymmetric amplitude  in
\reef{lop7399}, one needs to write down sort of new couplings, in the sense that in this turn, neither both scalar fields come from pull-back nor Taylor expansion.

Indeed for the first time, we just confirm the presence of sort of mixed couplings in string theory so that the first scalar field comes from pull-back and the second scalar comes through Taylor expansion. Thus the presence of mixed coupling  in EFT  is now being discovered. Note that this fact has become apparent by just dealing with $<V_{C^{-2}} V_{\phi^0}  V_{\phi^0}>$ S-matrix in an asymmetric picture of RR.

Let us write down the effective coupling at leading order which is $\alpha'^2$ and then generalize it to all orders in $\alpha'$. To be able to produce  \reef{lop7399}, at leading order one has to write down the following new coupling

\beqa S_6 = \frac{i(2\pi\alpha')^2\mu_p} {2p!}\int d^{p+1}\sigma
(\veps^v)^{a_0\cdots a_{p}} \bigg(\Tr( \Phi^i D_{a_{0}}\Phi^j)
\prt_i C^{(p+1)}_{ja_1\cdots a_{p}}+\Tr( \Phi^j D_{a_{0}}\Phi^i)
\prt_j C^{(p+1)}_{ia_1\cdots a_{p}} \bigg) \label{taylorpull} \eeqa

As it is clear from \reef{infinite}, we have many contact interactions which are related to higher derivative corrections of the scalar fields. These corrections without any ambiguities can be fixed by just S-matrix method. One can now apply the higher derivative corrections to scalar fields  to actually get to all order $\alpha'$ corrections of the above couplings as follows
 \beqa S_7  &=&i \frac{(2\pi\alpha')^2\mu_p} {2p!}\int d^{p+1}\sigma
(\veps^v)^{a_0\cdots a_{p}}\sum_{n=-1}^{\infty}c_{n}(\alpha')^{n+1} \nonumber\\&&\times
\bigg(\Tr( \prt_{a_{1}}...\prt_{a_{n+1}}\Phi^i D_{a_{0}}\prt^{a_{1}}...\prt^{a_{n+1}}\Phi^j)
\prt_i C^{(p+1)}_{ja_1\cdots a_{p}}\nonumber\\&&+\Tr( \prt_{a_{1}}...\prt_{a_{n+1}}\Phi^j  D_{a_{0}}\prt^{a_{1}}...\prt^{a_{n+1}}\Phi^i)
\prt_j C^{(p+1)}_{ia_1\cdots a_{p}}\bigg) \label{taylorpul2} \eeqa

Once more one could restore the gauge invariance  by replacing the covariant derivatives in \reef{taylorpul2}. Now let us deal with the five point function of an RR and three scalar fields in an asymmetric picture.

\section{ $<V_{C^{-2}} V_{\phi^0}  V_{\phi^0} V_{\phi^0}>$ amplitude }

The S-matrix elements of three transverse scalar fields and one RR in an asymmetric picture (C- field)
 $<V_{C^{-2}} V_{\phi^0}  V_{\phi^0} V_{\phi^0}>$ at disk level 
can be presented by the following correlation functions 
\begin{eqnarray}
{\cal A}^{C^{-2} \phi^0 \phi^0 \phi^0} & \sim & \int dx_{1}dx_{2}dx_{3}dzd\bar{z}\,
  \lan V_{\phi}^{(0)}{(x_{1})}
V_{\phi}^{(0)}{(x_{2})}V_{\phi}^{(0)}{(x_{3})}
V_{RR}^{(-\frac{3}{2},-\frac{1}{2})}(z,\bar{z})\ran,\labell{sstring333}\eeqa
One can extract the whole S-matrix and divide it out to various correlation functions.

\vskip 0.1in

In order to be able to explore all fermionic correlations including the correlations of various currents with spin operators, one has to reconsider the modified Wick-like rule as mentioned in the last section. Let us simplify the S-matrix further\footnote{where in type II we set $\alpha'=2$.}
\beqa
{\cal A}^{C^{-2}\phi ^{0}\phi^{0}\phi^{0}}&\sim&\int dx_{1}dx_{2} dx_{3}dx_{4}dx_{5}(P_{-}\fsC_{(n-1)}M_p)^{\al\be}I\xi_{1i}\xi_{2j}\xi_{3k}x_{45}^{-3/4}\nonumber\\&&\times
\bigg((x_{45}^{-5/4} C^{-1}_{\alpha\beta})\bigg[-\eta^{ij} x_{12}^{-2}a^k_3
-\eta^{ik} x_{13}^{-2}a^j_2
-\eta^{jk} x_{23}^{-2} a^i_1+a^i_1 a^j_2 a^k_3\bigg]+a^{ik}_{13}a^{ja}_2+a^{jk}_{23}a^{ic}_3\nonumber\\&&+a^k_3a^{jaic}_4
+a^{ij}_{12}a^{kb}_5+a^i_1a^{kbja}_6+a^j_2a^{kbic}_7-i\alpha'^3 k_{1c}k_{2a}k_{3b} I_8^{kbjaic}\bigg)\Tr(\lam_1\lam_2\lam_3)
\labell{amp3q333},\eeqa
where $x_{ij}=x_i-x_j$, and also
\beqa
I&=&|x_{12}|^{\alpha'^2 k_1.k_2}|x_{13}|^{\alpha'^2 k_1.k_3}|x_{14}x_{15}|^{\frac{\alpha'^2}{2} k_1.p}|x_{23}|^{\alpha'^2 k_2.k_3}|
x_{24}x_{25}|^{\frac{\alpha'^2}{2} k_2.p}
|x_{34}x_{35}|^{\frac{\alpha'^2}{2} k_3.p}|x_{45}|^{\frac{\alpha'^2}{4}p.D.p},\nonumber\\
a^i_1&=&ip^{i}\bigg(\frac{x_{54}}{x_{14}x_{15}}\bigg)\nonumber\\
a^{ij}_{12}&=&(-\eta^{ij} x_{12}^{-2}+a^i_1a^j_2)\nonumber\\
a^{ik}_{13}&=&(-\eta^{ik} x_{13}^{-2}+a^i_1a^k_3)\nonumber\\
a^{jk}_{23}&=&(-\eta^{jk} x_{23}^{-2}+a^j_2a^k_3)\nonumber\\
a^j_2&=&ip^{j}\bigg(\frac{x_{54}}{x_{24}x_{25}}\bigg)\nonumber\\
a^k_3&=&ip^{k}\bigg(\frac{x_{54}}{x_{34}x_{35}}\bigg)\nonumber\\
a^{ja}_2&=&\alpha' ik_{2a}2^{-1}x_{45}^{-1/4}(x_{24}x_{25})^{-1} \bigg\{(\Gamma^{ja}C^{-1})_{\alpha\beta}\bigg\} ,\nonumber\\
a^{ic}_3&=&\alpha' ik_{1c}2^{-1}x_{45}^{-1/4}(x_{14}x_{15})^{-1}\bigg\{(\Gamma^{ic}C^{-1})_{\alpha\beta}\bigg\} ,\nonumber\\
a^{jaic}_4&=&-\alpha'^2 k_{1c}k_{2a}2^{-2}x_{45}^{3/4}(x_{14}x_{15}x_{24}x_{25})^{-1}\bigg\{(\Gamma^{jaic}C^{-1})_{\alpha\beta}+\alpha' n_1\frac{Re[x_{14}x_{25}]}{x_{12}x_{45}}+\alpha'^2 n_2\bigg(\frac{Re[x_{14}x_{25}]}{x_{12}x_{45}}\bigg)^{2}\bigg\}
,\nonumber\\
a^{kb}_5&=&\alpha' ik_{3b}2^{-1}x_{45}^{-1/4}(x_{34}x_{35})^{-1}\bigg\{(\Gamma^{kb}C^{-1})_{\alpha\beta}\bigg\} ,\nonumber\\
a^{kbja}_6&=&-\alpha'^2 k_{2a}k_{3b}2^{-2}x_{45}^{3/4}(x_{34}x_{35}x_{24}x_{25})^{-1}\bigg\{(\Gamma^{kbja}C^{-1})_{\alpha\beta}+\alpha' n_3\frac{Re[x_{24}x_{35}]}{x_{23}x_{45}}+\alpha'^2 n_4\bigg(\frac{Re[x_{24}x_{35}]}{x_{23}x_{45}}\bigg)^{2}\bigg\}
,\nonumber\\
a^{kbic}_7&=&-\alpha'^2 k_{1c}k_{3b}2^{-2}x_{45}^{3/4}(x_{34}x_{35}x_{14}x_{15})^{-1}\bigg\{(\Gamma^{kbic}C^{-1})_{\alpha\beta}+\alpha' n_5\frac{Re[x_{14}x_{35}]}{x_{13}x_{45}}+\alpha'^2 n_6\bigg(\frac{Re[x_{14}x_{35}]}{x_{13}x_{45}}\bigg)^{2}\bigg\}
\nonumber\eeqa
where all $n_{i}'s $ as well as $I_8^{kbjaic}$ are given in the Appendix 1. One finds that the amplitude is SL(2,R) invariant so to remove the volume of conformal killing group as well as for practical reasons we fix the position of open strings at  zero, one  and infinity, that is
$ x_{1}=0 , x_{2}=1,x_{3}\rightarrow \infty$. Now if we come over the evaluation of all the integrals on the location of closed string RR on upper half plane as explained in Appendix 1, then one finally reads off  all the elements of the string amplitude in an asymmetric picture of RR as follows
\beqa {\cal A}^{C^{-2}\phi^{0} \phi^{0} \phi^{0} }&=&{\cal A}_{1}+{\cal A}_{2}+{\cal A}_{3}+{\cal A}_{41}+{\cal A}_{42}+{\cal A}_{43}+{\cal A}_{5}\nonumber\\&&{\cal A}_{61}+{\cal A}_{62}+{\cal A}_{63}+{\cal A}_{71}+{\cal A}_{72}+{\cal A}_{73}+{\cal A}_{81}+{\cal A}_{82}\nonumber\\&&{\cal A}_{83}+{\cal A}_{84}+{\cal A}_{85}+{\cal A}_{86}+{\cal A}_{87}+{\cal A}_{88}+{\cal A}_{89}+{\cal A}_{810}
\labell{711u}\eeqa
where
\beqa
{\cal A}_{1}&\!\!\!\sim\!\!\!&i\Tr(P_{-}\fsC_{(n-1)}M_p)
\bigg[p.\xi_1 p.\xi_2 p.\xi_3Q_1+\xi_{1}.\xi_{2} p.\xi_3Q_2+\xi_{1}.\xi_{3} p.\xi_2Q_3+\xi_{3}.\xi_{2} p.\xi_1Q_4\bigg],
\nonumber\\
{\cal A}_{2}&\sim&i k_{2a}\xi_{2j}\Tr(P_{-}\fsC_{(n-1)}M_p \Gamma^{ja})
\bigg\{-\xi_{1}.\xi_{3}Q_3-p.\xi_1 p.\xi_3Q_1\bigg\}\nonumber\\
{\cal A}_{3}&\sim&i k_{1c}\xi_{1i}\Tr(P_{-}\fsC_{(n-1)}M_p \Gamma^{ic})
\bigg\{-\xi_{2}.\xi_{3}Q_4-p.\xi_2 p.\xi_3Q_1\bigg\}\nonumber\\
{\cal A}_{41}&\sim&i\Tr(P_{-}\fsC_{(n-1)}M_p \Gamma^{jaic})\xi_{2j}\xi_{1i} p.\xi_3k_{1c}k_{2a}Q_1\nonumber\\
{\cal A}_{42}&\sim& ip.\xi_3 L_2\bigg\{t\xi_{1i}\xi_{2j}\Tr(P_{-}\fsC_{(n-1)}M_p \Gamma^{ji})-2\xi_{1}.\xi_{2}\Tr(P_{-}\fsC_{(n-1)}M_p \Gamma^{ac})k_{1c}k_{2a}\bigg\}
\nonumber\\
{\cal A}_{43}&\sim& 2itp.\xi_3 \xi_1.\xi_2 \Tr(P_{-}\fsC_{(n-1)}M_p )Q_5\bigg\{us-\frac{u+s+t}{2}\bigg\}\\
\nonumber\\
{\cal A}_{5}&\sim&i k_{3b}\xi_{3k}\Tr(P_{-}\fsC_{(n-1)}M_p \Gamma^{kb})
\bigg\{-\xi_{2}.\xi_{1}Q_2-p.\xi_2 p.\xi_1Q_1\bigg\}\nonumber\\
{\cal A}_{61}&\sim&i\Tr(P_{-}\fsC_{(n-1)}M_p \Gamma^{kbja})\xi_{2j}\xi_{3k} p.\xi_1k_{3b}k_{2a}Q_1\nonumber\\
{\cal A}_{62}&\sim& ip.\xi_1 L_5\bigg\{u\xi_{2j}\xi_{3k}\Tr(P_{-}\fsC_{(n-1)}M_p \Gamma^{kj})-2\xi_{2}.\xi_{3}\Tr(P_{-}\fsC_{(n-1)}M_p \Gamma^{ba})k_{2a}k_{3b}\bigg\}
\nonumber\\
{\cal A}_{63}&\sim& 2iup.\xi_1 \xi_2.\xi_3 \Tr(P_{-}\fsC_{(n-1)}M_p )Q_6\bigg\{st-\frac{u+s+t}{2}\bigg\}
\nonumber\\
{\cal A}_{71}&\sim&i\Tr(P_{-}\fsC_{(n-1)}M_p \Gamma^{kbic})\xi_{1i}\xi_{3k} p.\xi_2k_{3b}k_{1c}Q_1\nonumber\\
{\cal A}_{72}&\sim& ip.\xi_2 L_3\bigg\{-s\xi_{1i}\xi_{3k}\Tr(P_{-}\fsC_{(n-1)}M_p \Gamma^{ki})+2\xi_{1}.\xi_{3}\Tr(P_{-}\fsC_{(n-1)}M_p \Gamma^{bc})k_{1c}k_{3b}\bigg\}
\nonumber\\
{\cal A}_{73}&\sim& 2isp.\xi_2 \xi_1.\xi_3 \Tr(P_{-}\fsC_{(n-1)}M_p )Q_7\bigg\{ut-\frac{u+s+t}{2}\bigg\}\nonumber\\
{\cal A}_{81}&\sim&-i\Tr(P_{-}\fsC_{(n-1)}M_p \Gamma^{kbjaic})\xi_{1i}\xi_{2j}\xi_{3k}k_{1c} k_{2a}k_{3b}Q_1\nonumber\\
{\cal A}_{82}&\sim& -i L_2\bigg\{t\xi_{1i}\xi_{2j}\xi_{3k}k_{3b}\Tr(P_{-}\fsC_{(n-1)}M_p \Gamma^{kbji})-2\xi_{1}.\xi_{2}\Tr(P_{-}\fsC_{(n-1)}M_p \Gamma^{kbac})\xi_{3k}k_{1c}k_{2a}k_{3b}\bigg\}
\nonumber\\
{\cal A}_{83}&\sim& -i L_3\bigg\{-s\xi_{1i}\xi_{2j}\xi_{3k}k_{2a}\Tr(P_{-}\fsC_{(n-1)}M_p \Gamma^{kjai})+2\xi_{1}.\xi_{3}\Tr(P_{-}\fsC_{(n-1)}M_p \Gamma^{bjac})\xi_{2j}k_{1c}k_{2a}k_{3b}\bigg\}
\nonumber\\
{\cal A}_{84}&\sim& i L_5\bigg\{-u\xi_{1i}\xi_{2j}\xi_{3k}k_{1c}\Tr(P_{-}\fsC_{(n-1)}M_p \Gamma^{kjic})+2\xi_{2}.\xi_{3}\Tr(P_{-}\fsC_{(n-1)}M_p \Gamma^{baic})\xi_{1i}k_{1c}k_{2a}k_{3b}\bigg\}
\nonumber\\
{\cal A}_{85}&\sim& -2it \xi_{3k} k_{3b}\xi_1.\xi_2 \Tr(P_{-}\fsC_{(n-1)}M_p \Gamma^{kb})Q_5\bigg\{us-\frac{u+s+t}{2}\bigg\}\nonumber\\
{\cal A}_{86}&\sim& -iuL_6 \bigg\{2t\xi_1.\xi_3 k_{3b}\xi_{2j}\Tr(P_{-}\fsC_{(n-1)}M_p \Gamma^{bj})
-2s\xi_1.\xi_2 k_{2a}\xi_{3k}\Tr(P_{-}\fsC_{(n-1)}M_p \Gamma^{ka})\bigg\}\nonumber\\
{\cal A}_{87}&\sim& isL_6 \bigg\{-2t\xi_2.\xi_3 k_{3b}\xi_{1i}\Tr(P_{-}\fsC_{(n-1)}M_p \Gamma^{bi})
+2u\xi_1.\xi_2 k_{1c}\xi_{3k}\Tr(P_{-}\fsC_{(n-1)}M_p \Gamma^{kc})\bigg\}\nonumber\\
{\cal A}_{88}&\sim& -2is \xi_{2j} k_{2a}\xi_1.\xi_3 \Tr(P_{-}\fsC_{(n-1)}M_p \Gamma^{ja})Q_7\bigg\{ut-\frac{u+s+t}{2}\bigg\}
\nonumber\\
{\cal A}_{89}&\sim& -2iu \xi_{1i} k_{1c}\xi_2.\xi_3 \Tr(P_{-}\fsC_{(n-1)}M_p \Gamma^{ic})Q_6\bigg\{st-\frac{u+s+t}{2}\bigg\}
\nonumber\\
{\cal A}_{810}&\sim& -itL_6 \bigg\{2s\xi_2.\xi_3 k_{2a}\xi_{1i}\Tr(P_{-}\fsC_{(n-1)}M_p \Gamma^{ai})
-2u\xi_1.\xi_3 k_{1c}\xi_{2j}\Tr(P_{-}\fsC_{(n-1)}M_p \Gamma^{jc})\bigg\}
\labell{483333}\eeqa
where the functions
 $Q_1,Q_2,Q_3,Q_4,Q_5,Q_6,Q_7,L_2,L_3,L_5,L_6$ are given in Appendix 2.
Let us write down the same amplitude in its symmetric picture, that is,  $<V_{C^{-1}} V_{\phi^{-1}}  V_{\phi^0} V_{\phi^0}>$, which means that we consider the amplitude  in terms of field strength of RR and start comparing these two results. Working out in detail and making use of the integrals that presented in
\cite{Fotopoulos:2001pt} and \cite{Hatefi:2012wj},
 one constructs the S-matrix in symmetric picture of RR as follows
%%%
\beqa {\cal A}^{C^{-1}\phi^{-1}\phi^{0}\phi^{0} }&=&{\cal A'}_{1}+{\cal A'}_{2}+{\cal A'}_{3}+{\cal A'}_{4}+{\cal A'}_{5}+{\cal A'}_{6}
+{\cal A'}_{7}+{\cal A'}_{8}+{\cal A'}_{9}+{\cal A'}_{10}\labell{11u}
\eeqa
with
\beqa
{\cal A'}_{1}&\!\!\!\sim\!\!\!&-2^{-1/2}\xi_{1i}\xi_{2j}\xi_{3k}
\bigg[k_{3b}k_{2a}\Tr(P_{-}\fsH_{(n)}M_p\Gamma^{kbjai})-k_{3b}p^j\Tr(P_{-}\fsH_{(n)}M_p\Gamma^{kbi})\nonumber\\&&-k_{2a}p^k\Tr(P_{-}\fsH_{(n)}M_p\Gamma^{jai})+p^jp^k\Tr(P_{-}\fsH_{(n)}M_p\gamma^{i})\bigg]
Q_1,
\nonumber\\
{\cal A'}_{2}&\sim&2^{-1/2}
\bigg\{
2\xi_{1}.\xi_{2}k_{2a}k_{3b}\xi_{3k}\Tr(P_{-}\fsH_{(n)}M_p \Gamma^{kba})\bigg\}L_2\nonumber\\
{\cal A'}_{3}&\sim&2^{-1/2}
\bigg\{\xi_{1i}\xi_{2j}\xi_{3k}\Tr(P_{-}\fsH_{(n)}M_p \Gamma^{kji})\bigg\}(-uL_{5})\nonumber\\
{\cal A'}_{4}&\sim&-2^{-1/2}
\bigg\{
2\xi_{3}.\xi_{1}k_{2a}k_{3b}\xi_{2j}\Tr(P_{-}\fsH_{(n)}M_p \Gamma^{bja})\bigg\}L_3\nonumber\\
\nonumber\\
{\cal A'}_{5}&\sim&2^{-1/2}
\bigg\{2\xi_{3}.\xi_{2}k_{2a}k_{3b}\xi_{1i}\Tr(P_{-}\fsH_{(n)}M_p \Gamma^{bai})\bigg\}L_5\nonumber\\
{\cal A'}_{6}&\sim&2^{1/2}L_{2}\bigg\{-k_{2a}p^k\xi_1.\xi_2\xi_{3k}\Tr(P_{-}\fsH_{(n)}M_p\gamma^a)
\bigg\}
\nonumber\\
{\cal A'}_{7}&\sim&2^{1/2}L_3\bigg\{
k_{3b}p^j\xi_1.\xi_3\xi_{2j}\Tr(P_{-}\fsH_{(n)}M_p\gamma^b)\bigg\}%\delta_{p,n-1}
\nonumber\\
{\cal A'}_{8}&\sim&2^{1/2}L_6\bigg\{\xi_{2j}\Tr(P_{-}\fsH_{(n)}M_p\gamma^j)
(ut\xi_1.\xi_3)\bigg\}.
\nonumber\\
{\cal A'}_{9}&\sim&2^{1/2}L_6\bigg\{\xi_{3k}\Tr(P_{-}\fsH_{(n)}M_p\gamma^k)
(us\xi_1.\xi_2)\bigg\}
\nonumber\\
{\cal A'}_{10}&\sim&2^{1/2}L_6\bigg\{\xi_{1i}\Tr(P_{-}\fsH_{(n)}M_p\gamma^i)
(ts\xi_3.\xi_2)\bigg\}
\labell{480}
\eeqa
%%%
where  $L_2,L_3,L_5,L_6$ are introduced in \reef{Ls}. Let us compare the amplitudes in asymmetric picture with its symmetric result and explore all new bulk singularity structures 
as well as  new couplings (with their all order $\alpha'$ corrections) that are just appeared in asymmetric picture of the amplitude.
%%%
%\beqa
%L_1&=&(2)^{-2(t+s+u)+1}\pi{\frac{\Gamma(-u+\frac{1}{2})
%\Gamma(-s+\frac{1}{2})\Gamma(-t+\frac{1}{2})\Gamma(-t-s-u+1)}
%{\Gamma(-u-t+1)\Gamma(-t-s+1)\Gamma(-s-u+1)}},\nonumber\\
%L_2&=&(2)^{-2(t+s+u)}\pi{\frac{\Gamma(-u+1)
%\Gamma(-s+1)\Gamma(-t)\Gamma(-t-s-u+\frac{1}{2})}
%{\Gamma(-u-t+1)\Gamma(-t-s+1)\Gamma(-s-u+1)}}
%\nonumber\\
%L_{22}&=&(2)^{-2(t+s+u)}\pi{\frac{\Gamma(-u+1)
%\Gamma(-s+1)\Gamma(-t+1)\Gamma(-t-s-u+\frac{1}{2})}
%{\Gamma(-u-t+1)\Gamma(-t-s+1)\Gamma(-s-u+1)}}
%\nonumber\\
%L_{33}&=&(2)^{-2(t+s+u)}\pi{\frac{\Gamma(-u+1)
%\Gamma(-s+1)\Gamma(-t)\Gamma(-t-s-u+\frac{1}{2})}{\Gamma(-u-t+1)
%\Gamma(-t-s+1)\Gamma(-s-u+1)}}
%\nonumber\\
%L_3&=&(2)^{-2(t+s+u)}\pi{\frac{\Gamma(-u+1)
%\Gamma(-s)\Gamma(-t+1)\Gamma(-t-s-u+\frac{1}{2})}{\Gamma(-u-t+1)
%\Gamma(-t-s+1)\Gamma(-s-u+1)}}
%,\nonumber\\
%L_4&=&(2)^{-2(t+s+u)}\pi{\frac{\Gamma(-u+1)
%\Gamma(-s)\Gamma(-t+1)\Gamma(-t-s-u+\frac{1}{2})}
%{\Gamma(-u-t+1)\Gamma(-t-s+1)\Gamma(-s-u+1)}}
%,\nonumber\\
%L_5&=&(2)^{-2(t+s+u)}\pi{\frac{\Gamma(-u)
%\Gamma(-s+1)\Gamma(-t+1)\Gamma(-t-s-u+\frac{1}{2})}
%{\Gamma(-u-t+1)\Gamma(-t-s+1)\Gamma(-s-u+1)}}
%,\nonumber\\
%L_6&=&(2)^{-2(t+s+u)-1}\pi{\frac{\Gamma(-u+\frac{1}{2})
%\Gamma(-s+\frac{1}{2})\Gamma(-t+\frac{1}{2})\Gamma(-t-s-u)}
%{\Gamma(-u-t+1)\Gamma(-t-s+1)\Gamma(-s-u+1)}},
%\label{Ls}
%\eeqa
%%%

\section{Singularity Comparisons between asymmetric and symmetric pictures}

In this section we try to  produce all the singularity structures of symmetric picture by dealing with the S-matrix element in  asymmetric picture of RR. We add  the 1st term of ${\cal A}_{810}$ of asymmetric amplitude with first term of ${\cal A}_{87}$ and apply momentum conservation along the world volume of brane to actually reach to the following couplings
\beqa
 -2istL_6 \xi_2.\xi_3 \xi_{1i}\Tr(P_{-}\fsC_{(n-1)}M_p \Gamma^{bi})(k_{3b}+k_{2b})=-2istL_6 \xi_2.\xi_3 \xi_{1i}\Tr(P_{-}\fsC_{(n-1)}M_p \Gamma^{bi})(-k_{1b}-p_{b})\nonumber\eeqa

now if we use  $p\fsC =\fsH$ then we see that above coupling precisely generates ${\cal A'}_{10}$ of symmetric picture ($<V_{C^{-1}} V_{\phi^{-1}}V_{\phi^{0}}V_{\phi^{0}}>$) \footnote{up to a normalisation constant $2^{1/2}i$}. The first term of above equation will be precisely cancelled off  by sum of  the first term of ${\cal A}_{3}$ of asymmetric picture and the whole ${\cal A}_{89}$.

Likewise what we did previously, here we try to add  the second terms of ${\cal A}_{86}$ and  ${\cal A}_{87}$  and in particular we take into account  the momentum conservation to actually arrive at the following singularities
\beqa
 2isuL_6 \xi_2.\xi_1 \xi_{3k}\Tr(P_{-}\fsC_{(n-1)}M_p \Gamma^{kc})(-k_{3c}-p_{c})\label{nmo8101}\eeqa

Using $p\fsC =\fsH$ we are able to reconstruct  ${\cal A'}_{9}$ of $<V_{C^{-1}} V_{\phi^{-1}}V_{\phi^{0}}V_{\phi^{0}}>$. The first term of \reef{nmo8101} will be removed by the sum of the first term of ${\cal A}_{5}$  and the whole ${\cal A}_{85}$ of asymmetric S-matrix . The same holds as follows.

\vskip.2in
Adding  the 2nd term of ${\cal A}_{810}$ and the first term of ${\cal A}_{86}$  also paying particular attention to momentum conservation give rise to the following elements
\beqa
 -2ituL_6 \xi_3.\xi_1 \xi_{2j}\Tr(P_{-}\fsC_{(n-1)}M_p \Gamma^{bj})(-k_{2b}-p_{b})\label{nmo8102}\eeqa

which is precisely  ${\cal A'}_{8}$. Notice to the point  that the first term of \reef{nmo8102} has been  equivalently cancelled off by  the sum of  first term of ${\cal A}_{2}$ and the entire ${\cal A}_{88}$ of $<V_{C^{-2}} V_{\phi^{0}}V_{\phi^{0}}V_{\phi^{0}}>$.

Having taken the second term of ${\cal A}_{72}$ and having applied the momentum conservation to it we seem to get
\beqa
2ip.\xi_2 \xi_3.\xi_1   L_3\Tr(P_{-}\fsC_{(n-1)}M_p \Gamma^{bc})k_{3b} (-k_{3c}-p_{c}-k_{2c})\label{nmo8103}\eeqa
evidently  the first term above has no contribution, because there is an antisymmetric $\epsilon$ tensor while the whole element is symmetric with respect to interchanging $k_3$ so the answer turns out to be zero while  the second term in  \reef{nmo8103} precisely generates ${\cal A'}_{7}$. The the last term in \reef{nmo8103}  remains to be explored later on.

By applying the same tricks to the second term of ${\cal A}_{42}$ we obtain
\beqa
-2ip.\xi_3 \xi_2.\xi_1   L_2\Tr(P_{-}\fsC_{(n-1)}M_p \Gamma^{ac})k_{2a} (-k_{2c}-p_{c}-k_{3c})\label{nmo8104}\eeqa
obviously the first term above has zero contribution to S-Matrix and the second term reconstructs ${\cal A'}_{6}$, meanwhile  the last term will be considered in the next section. We also need to  take into account the 2nd term of ${\cal A}_{84}$ and apply the momentum conservation to it to get to
\beqa
2iL_5\xi_2.\xi_3   \Tr(P_{-}\fsC_{(n-1)}M_p \Gamma^{baic}) \xi_{1i}  k_{2a}  k_{3b}(-k_{2c}-k_{3c}-p_{c})\label{nmo8105}\eeqa
The first and second term in \reef{nmo8105} have zero contribution and the third  term produces ${\cal A'}_{5}$. Once more one needs to  deal with the 2nd term of ${\cal A}_{83}$ and draw attention to  momentum conservation in such a way that the following singularities turn out to be produced
\beqa
-2iL_3\xi_1.\xi_3   \Tr(P_{-}\fsC_{(n-1)}M_p \Gamma^{bjac}) \xi_{2j}  k_{2a}  k_{3b}(-k_{2c}-k_{3c}-p_{c})\label{nmo8106}\eeqa

clearly the first and second term of \reef{nmo8106} have no contribution to amplitude as there is an antisymmetric $\epsilon$ tensor while the whole singularity is symmetric with respect to interchanging $k_2,k_3$ so the answer turns out to be zero while the third term in above singularity re-builds ${\cal A'}_{4}$.

\vskip.1in

We need to  carry out the same tricks to the second term of ${\cal A}_{82}$ to be able to recreate
exactly  ${\cal A'}_{2}$ of symmetric amplitude.

\vskip.2in

Eventually we need to  add the 1st  terms of ${\cal A}_{82}$, ${\cal A}_{83}$ and ${\cal A}_{84}$ to actually derive the following singularities
\beqa
iL_{22}\xi_{1i}\xi_{2j} \xi_{3k}  \Tr(P_{-}\fsC_{(n-1)}M_p \Gamma^{kbji}) (k_{1b}+k_{2b}+k_{3b})=iL_{22}\xi_{1i}\xi_{2j} \xi_{3k}  \Tr(P_{-}\fsC_{(n-1)}M_p \Gamma^{kbji}) (-p_b)\nonumber\eeqa
which is nothing but exactly ${\cal A'}_{3}$ part of symmetric S-matrix and $L_{22}=-u L_5$.From the above comparisons we come to the following points. 

By doing careful analysis of asymmetric elements  we were able to re-generate  all order   $\alpha'$  singularity structures of symmetric amplitude ($<V_{C^{-1}}V_{\phi^{-1}} V_{\phi^{0}}V_{\phi^{0}}>$). However, the important point that must be emphasized is as follows. Regarding our true comparisons and the remarks that have already been pointed out in this section, we have got some extra contact interactions as well as two extra kinds of bulk singularity structures in asymmetric amplitude that can not be shown up by symmetric analysis and all of them will be highlighted in the next section. We also try to introduce new couplings in an EFT to be able to produce all those new bulk singularities as well.

\subsection{ Bulk singularity structures in Asymmetric Picture  }

\vskip.1in

As we have already argued,  the S-matrix of an RR and three transverse scalars  in its asymmetric picture (in addition to all the singularities of symmetric picture) generates two different kinds of bulk singularity structures.  For instance
%  poles in this asymmetric picture which are absent in symmetric picture, of $C^{-1}\phi^{-1}\phi^{0}\phi^{0}$,
 the second term of ${\cal A}_{62}$ of asymmetric picture ($<V_{C^{-2}}V_{\phi^{0}} V_{\phi^{0}}V_{\phi^{0}}>$) has got a new kind of infinite u- channel bulk singularities which can not be obtained from
the symmetric picture of $<V_{C^{-1}}V_{\phi^{-1}} V_{\phi^{0}}V_{\phi^{0}}>$. Indeed we  would have expected to have these bulk singularities in asymmetric picture, because of the symmetries  with respect to interchanging all three scalars as well as symmetries of string  amplitude.

Therefore let us point out the first  kind  of u-channel bulk singularity structure  in asymmetric analysis (which is the 2nd term of ${\cal A}_{62}$) of ($<V_{C^{-2}}V_{\phi^{0}} V_{\phi^{0}}V_{\phi^{0}}>$) as follows

\beqa
 -2ip.\xi_1 L_5 \xi_2.\xi_3 \Tr(P_{-}\fsC_{(n-1)}M_p \Gamma^{ba}) k_{3b} k_{2a}\label{newsins1}\eeqa

as well as  $t,s$- channel bulk singularities  coming from the 2nd terms of ${\cal A}_{42}$  or \reef{nmo8104} 
and ${\cal A}_{72}$ or \reef{nmo8103} appropriately as follows

 \beqa
 2ip.\xi_3 L_2 \xi_2.\xi_1 \Tr(P_{-}\fsC_{(n-1)}M_p \Gamma^{ac}) k_{3c} k_{2a}\nonumber\\
 -2ip.\xi_2 L_3 \xi_1.\xi_3 \Tr(P_{-}\fsC_{(n-1)}M_p \Gamma^{bc}) k_{3b} k_{2c}\label{newsins}\eeqa

where in the first and second  equations one can use momentum conservation to actually write  $k_{3c}$ ($k_{2c}$) in terms of $k_{1c}$. Regarding the symmetries of S-matrix we just produce u-channel bulk singularities then by interchanging momenta and polarizations we can easily produce $t,s$- channel bulk singularities as well.

\vskip.2in

Let us first generate these new u-channel bulk singularities of the S-matrix. We need to consider the following rule
\beqa {\cal
A}&=&V^a_{\alpha}(C_{p-1},\phi_1,
A)G^{ab}_{\alpha\beta}(A)V^b_{\beta}(A,\phi_2,\phi_3),\labell{ampvv1}
\eeqa
The kinetic terms of scalar field and gauge field have been taken into account to reach at the following vertex and propagator
\beqa
V_\beta^{b}(A,\phi_2,\phi_3)&=&iT_p(2\pi\alpha')^{2}
\xi_{2}.\xi_{3}(k_2-k_3)^{b}\Tr(\lambda_{2}\lambda_{3}\lambda_{\beta}),
\nonumber\\
G_{\alpha\beta}^{ab}(A)&=&\frac{i\delta_{\alpha\beta}\delta^{ab}}{(2\pi\alpha')^{2}T_p u }\label{oop1}\eeqa

In order to find out $V^a_{\alpha}(C_{p-1},\phi_1,A)$ , one has to employ Taylor expansion as follows
\beqa S_8
&=&{i}(2\pi\alpha')^2\mu_p\int d^{p+1}\s {1\over (p-1)!}(\veps^v)^{a_0\cdots a_{p}}
\left[ \partial_{i}C_{ a_{0}... a_{p-2}}F_{a_{p-1}a_{p}}\phi^i
\right]\,\,\,
\label{fin1781}
\eeqa

then take the integration by parts. The gauge field here is Abelian, taking \reef{fin1781} to momentum space and consider the following equation
\beqa
-p.\xi_1 C_{ a_{0}... a_{p-2}} {(p+k_1)}_{a_{p-1}}&=& p.\xi_1 C_{ a_{0}... a_{p-2}} {(k_2+k_3)}_{a_{p-1}}
\nonumber\eeqa
 we can obtain
\beqa
V^a_{\alpha}(C_{p-1},\phi_1,A)&=& {i}(2\pi\alpha')^2\mu_p(\veps^v)^{a_0\cdots a_{p-1}a} p.\xi_1 C_{ a_{0}... a_{p-2}} {(k_2+k_3)}_{a_{p-1}}\Tr(\lambda_{1}\lambda_{\alpha})
\label{esi98o}\eeqa

 Both propagator and $V_\beta^{b}(A,\phi_2,\phi_3)$ are derived from kinetic terms thus  there is no any correction to these terms. Therefore in order to explore all infinite u-channel bulk poles one should impose all infinite higher derivative correction to \reef{fin1781} as 

 \beqa S_9
&=&{i}(2\pi\alpha')^2\mu_p\int d^{p+1}\s {1\over (p-1)!}(\veps^v)^{a_0\cdots a_{p}}\sum_{n=-1}^{\infty}b_n
(\alpha')^{n+1}\nonumber\\&&\times
\left[ \partial_{i}C_{ a_{0}... a_{p-2}}
D_{a_{0}}\cdots D_{a_{n}}F_{a_{p-1}a_{p}}D^{a_{0}}\cdots
D^{a_{n}}\phi^i
\right]\,\,\,
\label{fin1782}
\eeqa
%%%
to get to all order extension of the above vertex operator as
\beqa
V^a_{\alpha}(C_{p-1},\phi_1,A)&=&\sum_{n=-1}^{\infty}b_n
(\alpha')^{n+1}(k_1.k)^{n+1} {i}(2\pi\alpha')^2\mu_p(\veps^v)^{a_0\cdots a_{p-1}a}\nonumber\\&&\times
 p.\xi_1 C_{ a_{0}... a_{p-2}} {(k_2+k_3)}_{a_{p-1}}\Tr(\lambda_{1}\lambda_{\alpha})
\label{esiccx}\eeqa

Now if we replace \reef{esiccx} and \reef{oop1} inside \reef{ampvv1} then we are able to precisely produce all infinite  u-channel bulk singularities of  this amplitude as follows

\beqa
 -2ip.\xi_1 \frac{16 \mu_p\pi^2 }{u (p-1)!} \sum_{n=-1}^{\infty}b_n
(t+s)^{n+1}  \xi_2.\xi_3  k_{3b} k_{2a} C_{ a_{0}... a_{p-2}} (\veps^v)^{a_0\cdots a_{p-2}ba}
\eeqa

The second kind of new bulk singularity structure is as follows.

\vskip.3in

We consider  the following bulk singularity structures of asymmetric picture as well  which can be eventually simplified, by various
algebraic calculations as follows.

Indeed If we  add the 2nd term of ${\cal A}_{1}$ and the entire ${\cal A}_{43}$ of $<V_{C^{-2}} V_{\phi^{0}}V_{\phi^{0}}V_{\phi^{0}}>$ we get to derive
\beqa
 &&\Tr(P_{-}\fsC_{(n-1)}M_p)ip.\xi_3  \xi_2.\xi_1 \bigg(Q_2 +2t Q_5\bigg\{us-\frac{u+s+t}{2}\bigg\}\bigg)=\nonumber\\&&
 \Tr(P_{-}\fsC_{(n-1)}M_p) \bigg(-2ius p.\xi_3  \xi_2.\xi_1 \bigg)L_6 \label{sings12}\eeqa

 Likewise if we add up the 3rd term of ${\cal A}_{1}$ and the entire ${\cal A}_{73}$ of asymmetric picture we get to obtain
\beqa
 && \Tr(P_{-}\fsC_{(n-1)}M_p)  ip.\xi_2  \xi_3.\xi_1 \bigg(Q_3 +2s Q_7\bigg\{ut-\frac{u+s+t}{2}\bigg\}\bigg)=\nonumber\\&&
 \Tr(P_{-}\fsC_{(n-1)}M_p)\bigg(-2iut p.\xi_2  \xi_3.\xi_1 \bigg)L_6 \label{sings13}\eeqa

Finally we must add up  the 4th term of ${\cal A}_{1}$ and the whole ${\cal A}_{63}$ of  asymmetric S-matrix to be able to gain
\beqa
 &&   \Tr(P_{-}\fsC_{(n-1)}M_p) ip.\xi_1  \xi_3.\xi_2 \bigg(Q_4 +2u Q_6\bigg\{st-\frac{u+s+t}{2}\bigg\}\bigg)=\nonumber\\&&
\Tr(P_{-}\fsC_{(n-1)}M_p)\bigg(-2ist p.\xi_1  \xi_3.\xi_2 \bigg) L_6 \label{sings14}\eeqa

 All the above terms must be added up as follows

\beqa
 \Tr(P_{-}\fsC_{(n-1)}M_p) p^i \bigg(-2ius \xi_{3i}  \xi_2.\xi_1 -2iut \xi_{2i}  \xi_3.\xi_1 -2ist \xi_{1i}  \xi_3.\xi_2 \bigg) L_6 \label{sings18}\eeqa

Having extracted the trace and replacing $L_6$ expansion in string amplitude, one finds out the second kind of  new bulk singularity structure of BPS branes as follows

 \beqa
&&16\pi\mu_p\frac{1}{(p+1)!(s+t+u)}\Tr(\lam_1\lam_2\lam_3)
\sum_{n,m=0}^{\infty}(a_{n,m}+b_{n,m})\bigg(-2ius\xi_{3i}\xi_1.\xi_2
[s^{m}u^{n}+s^{n}u^{m}]\nonumber\\&&-2iut\xi_{2i}\xi_1.\xi_3
[t^{m}u^{n}+t^{n}u^{m}]-2its\xi_{1i}\xi_3.\xi_2
[s^{m}t^{n}+s^{n}t^{m}]\bigg)\eps^{a_{0}\cdots a_{p}} p^i C_{a_0\cdots
a_{p}} \label{amphigh87} \eeqa
%%%

This second new bulk singularity structure is related to new structures of an infinite $(t+s+u)$ bulk singularities. Indeed $L_6$ does have infinite $(t+s+u)$ channel singularities and in order to produce them in an EFT one has to consider the following rule

\beqa
{\cal A}&=&V_{\alpha}^{i}(C_{p+1},\phi)G_{\alpha\beta}^{ij}(\phi)V_{\beta}^{j}(\phi,\phi_1,
\phi_2,\phi_3),\labell{amp5498}
\eeqa

The following coupling in an EFT is needed

\beqa
S_{10}&=&i(2\pi\alpha')\mu_p\int d^{p+1}\sigma {1\over (p)!}
(\veps^v)^{a_0\cdots a_{p}}\,
D_{a_{0}} \phi^i C^{(p+1)}_{ia_1\cdots a_{p}}
\label{298}\eeqa

where in \reef{298} the scalar field  has been taken from pull-back. The trace in \reef{sings18} shows the RR potential has to be $p+1$ form field. \reef{298} is  a new coupling that plays the crucial rule for matching all the infinite new bulk singularity structures of string amplitude with effective field theory. If we take into account \reef{298} and $p_{{a_{0}}} C^i_{a_1\cdots a_{p}}=p^i C_{a_0\cdots a_{p}} $ as well as the kinetic term of the scalar field $((2\pi\alpha')^2/2) D_a\phi^i D^a\phi_i$ we obtain
\beqa
V_{\alpha}^{i}(C_{p+1},\phi)&=&i(2\pi\alpha')\mu_p\frac{1}{(p)!}(\veps^v)^{a_0\cdots a_{p}}
 p^i C_{a_0\cdots a_{p}}\Tr(\lambda_{\alpha})\nonumber\\
G_{\alpha\beta}^{ij}(\phi) &=&\frac{-i\delta_{\alpha\beta}\delta^{ij}}{T_p(2\pi\alpha')^2
(t+s+u)}
\labell{Fey4}
\eeqa

One needs to also impose the infinite higher derivative corrections to four real scalar field couplings that are derived in \cite{Hatefi:2012ve} as
\beqa
(2\pi\alpha')^4\frac{1}{4\pi^2}T_p\left(\alpha'\right)^{n+m}\sum_{m,n=0}^{\infty}(\cL_{11}^{nm}+\cL_{12}^{nm}+\cL_{13}^{nm})\labell{lagrang1}\eeqa
\beqa &&\cL_{11}^{nm}=-
\Tr\left(\frac{}{}a_{n,m}\cD_{nm}[D_{a}\phi^i D_{b}\phi_i D^{b}\phi^j D^{a}\phi_j]+\frac{}{} b_{n,m}\cD'_{nm}[D_{a}\phi^i D^{b}\phi^j  D_{b}\phi_i
D^{a}\phi_j ]+h.c.\frac{}{}\right)\nonumber\\
&&\cL_{12}^{nm}=-\Tr\left(\frac{}{}a_{n,m}\cD_{nm}[D_{a}\phi^i D_{b}\phi_i D^{a}\phi^j D^{b}\phi_j]+\frac{}{}b_{n,m}\cD'_{nm}[D_{b}\phi^i D^{b}\phi^j
 D_{a}\phi_i  D^{a}\phi_j  ]+h.c.\frac{}{}\right)\nonumber\\
&&\cL_{13}^{nm}=\Tr\left(\frac{}{}a_{n,m}\cD_{nm}[D_{a}\phi^i D^{a}\phi_i D_{b}\phi^j D^{b}\phi_j]+\frac{}{}b_{n,m}\cD'_{nm}[D_{a}\phi^i D_{b}\phi^j
 D^{a}\phi_i  D^{b}\phi_j ]+h.c\frac{}{}\right)\nonumber
\eeqa

to actually obtain the following vertex
\beqa
V_{\beta}^{j}(\phi,\phi_1, \phi_2,\phi_3)
 &=&\frac{1}{4\pi^2}(\alpha')^{n+m}(a_{n,m}+b_{n,m})(2\pi\alpha')^4T_{p}
 \bigg(2ts \xi_{1}^j\xi_2.\xi_3 (s^{m}t^{n}+s^{n}t^{m})\nonumber\\&&
 +
 2ut \xi_{2}^j\xi_1.\xi_3 (t^{m}u^{n}+t^{n}u^{m})+ 2 us \xi_{3}^j\xi_1.\xi_2 (s^{m}u^{n}+s^{n}u^{m})\bigg)\nonumber\\&&\times
 \Tr(\lam_1\lam_2\lam_3\lambda_{\beta})  \label{pol900}\eeqa

Now if we replace \reef{pol900},\reef{Fey4} inside \reef{amp5498} then we are exactly able to produce all order $(t+s+u)$ channel bulk singularity structures  of an RR and three scalar fields  
 of string amplitude  \reef{amphigh87} in effective field theory side as well. Note that the important point was to derive the new coupling of a scalar field and a $p+1$ potential RR field as explored in \reef{298}.

\vskip.1in

Note that there had been another kind of $(t+s+u)$ channel poles which have already  been discovered in \cite{Hatefi:2012rx}.
%\beqa
%(2\pi\alpha')\mu_p\int d^{p+1}\sigma {1\over (p+1)!}
%(\veps^v)^{a_0\cdots a_{p}}\,\Tr\left(\phi^i\right)\,
%H_{ia_0\cdots a_{p}}\eeqa
For the completeness let us very briefly just produce all infinite u and t-channel poles accordingly through Field strength of RR. Eventually one can exchange the momenta and polarisations to get to all infinite  s-channel poles as well.
 \beqa
{\cal A}_{u}&=&\frac{32}{p!} \pi^{2}\mu_p
   \sum_{n=-1}^{\infty}\frac{1}{u}{b_n(t+s)^{n+1}} \xi_{2}.\xi_{3}k_{2a}k_{3b}\xi_{1k} \eps^{a_{0}\cdots a_{p-2}ba}H^{k}_{a_{0}\cdots a_{p-2}}
 \nonumber\eeqa

 Now if we impose the mixed couplings of RR's field strength, a gauge field and a scalar field through pull back as follows
 \beqa S_{11}&=& (2\pi\alpha')^2\mu_p\int d^{p+1}\s {1\over (p-1)!}(\veps^v)^{a_0\cdots a_{p}}
F_{a_{0}a_{1}}D_{a_{2}}\phi^k
 C^{(p-1)}_{k a_{3}... a_{p}}.
\label{fin1}
\eeqa
 and if we take integration by parts on scalar field we come to know that $D_{a_{2}}$ can just act on C-field, because of the antisymmetric property of  $(\veps^v)$ it cannot act on F. Thus one can explore the following vertex operator
 \beqa
V^a_{\alpha}(C_{p-1},\phi_1,A)&=&\frac{(2\pi\alpha')^2\mu_p}{(p)!}(\veps^v)^{a_0\cdots a_{p-1}a}(H^{(p)})^k{}_{a_0\cdots a_{p-2}}\xi_{1k}k_{a_{p-1}}\Tr(\lam_1\lambda_\alpha)\nonumber
\eeqa

$k$ is the momentum of off-shell gauge field $k=k_2+k_3$ and one imposes higher derivative corrections \footnote{\beqa S_{12}&=&{i}(2\pi\alpha')^2\mu_p\int d^{p+1}\s {1\over
(p-1)!}(\veps^v)^{a_0\cdots a_{p}} \nonumber \\&&\times
\sum_{n=-1}^{\infty}
b_n(\alpha')^{n+1}\left[\Tr\bigg(\partial_{a_{m_{0}}}\cdots
\partial_{a_{m_{n}}} F_{a_{0}a_{1}}\partial^{a_{m_{0}}}\cdots
\partial^{a_{m_{n}}}D_{a_{2}}\phi^k\bigg)
  C^{(p-1)}_{k a_{3}... a_{p}}\right]
\nonumber
\eeqa}
to be able to  get to  all order extension of vertex as follows
\beqa
V^a_{\alpha}(C_{p-1},\phi_1,A)&=&\frac{(2\pi\alpha')^2\mu_p}{(p)!}(\veps^v)^{a_0\cdots a_{p-1}a}\nonumber\\&&\times
H^k{}_{a_0\cdots a_{p-2}}\xi_{1k}k_{a_{p-1}}\Tr(\lam_1\lambda_\alpha)\sum_{n=-1}^{\infty}b_n(t+s)^{n+1},\label{mash}
\eeqa

Implementing \reef{mash} and \reef{oop1} inside \reef{ampvv1} one can exactly generate these infinite u-channel poles.
%(-k_2-k_3)^a(k2+k_3)^b\delta^{ab}=2k_2a k_3b
Eventually one reads off all infinite t-channel poles as follows
\beqa
{\cal A}_{2}&=&\pm  \frac{16}{p!} \pi^{2}\mu_p \sum_{n=-1}^{\infty}\frac{1}{t}{b_n(u+s)^{n+1}}\bigg(\eps^{a_{0}\cdots a_{p-2}ba}H^{k}_{a_{0}\cdots a_{p-2}} (2\xi_{1}.\xi_{2}k_{2a}k_{3b}\xi_{3k})\nonumber\\&&+p^k \eps^{a_{0}\cdots a_{p-1}a}H_{a_{0}\cdots a_{p-1}} (-2k_{2a}\xi_1.\xi_2\xi_{3k})\bigg)\Tr(\lam_1\lam_2\lam_3)\label{esi56}\eeqa
Having considered Taylor expansion for the scalar field
\beqa S_{13}&=& (2\pi\alpha')^2\mu_p\int d^{p+1}\s {1\over (p-1)!}(\veps^v)^{a_0\cdots a_{p}}
F_{a_{0}a_{1}}\phi^k
 \partial_{k}C^{(p-1)}_{ a_{2}... a_{p}}
\label{fin17r8}\eeqa
and taken integration by parts on the location of gauge field, we got the following action
\beqa
S_{13} &=& i(2\pi\alpha')^2\mu_p\int d^{p+1}\s
{1\over (p-1)!}(\veps^v)^{a_0\cdots a_{p}}
\bigg(-A_{a_{1}}\partial_{a_{0}}\phi^k
 \partial_{k}C^{(p-1)}_{ a_{2}... a_{p}} -A_{a_{1}}\phi^k \partial_{a_{0}} \partial_{k}C^{(p-1)}_{ a_{2}... a_{p}}\bigg)
\nonumber
\eeqa

We now apply the higher derivative  corrections to \reef{fin17r8} so that the following vertex can be achieved
\beqa
V^a_{\alpha}(C_{p-1},\phi_3,A)&=&\frac{(2\pi\alpha')^2\mu_p}{(p-1)!}(\veps^v)^{a_0\cdots
a_{p-1}a}\Tr(\lam_3\lambda_\alpha)\sum_{n=-1}^{\infty}b_n(s+u)^{n+1}
\nonumber\\&&\times  p^k \xi_{3k}(p+k_3)_{a_{p-1}}(C^{(p-1)})_{a_0\cdots a_{p-2}},
\label{mnb}\eeqa
Taking the rule as $
A =V^a_{\alpha}(C_{p-1},\phi_3,
A)G^{ab}_{\alpha\beta}(A)V^b_{\beta}(A,\phi_1,\phi_2)$
and the vertices as
\beqa
V_\beta^{b}(A,\phi_1,\phi_2)&=&iT_p(2\pi\alpha')^{2}
\xi_{1}.\xi_{2}(k_1-k_2)^{b}\Tr(\lambda_{1}\lambda_{2}\lambda_{\beta}),
\nonumber\\
G_{\alpha\beta}^{ab}(A)&=&\frac{i\delta_{\alpha\beta}\delta^{ab}}{(2\pi\alpha')^{2}T_p t},\nonumber\eeqa
 and also considering momentum conservation we obtain
\beqa
V_\beta^{b}(A,\phi_1,\phi_2)&=&iT_p(2\pi\alpha')^{2} \xi_{1}.\xi_{2}(-2k_2-k_3-p)^{b}\Tr(\lambda_{1}\lambda_{2}\lambda_{\beta})
\label{987b}
\eeqa

Now replacing  \reef{987b} and \reef{mnb} inside the above rule  and making use of the fact that $(k_3+p)^2=(k_1+k_2)^2=t$  we are exactly able to produce all infinite t-channel poles of \reef{esi56}.
%t/t=1= contact terms
Exchanging the momenta and polarizations we are also able to construct all infinite  s-channel poles as well. Let us discuss contact interactions.
  \subsection{ All order $\alpha'$ contact interactions  of asymmetric S-Matrix }

 To begin with, we start generating the contact terms of symmetric picture by comparing them with contact terms of asymmetric picture. We address other contact interactions that are appeared just in asymmetric S-Matrix and essentially we write down  new EFT couplings  and explore their all order $\alpha'$ higher derivative corrections accordingly. First let us apply the the momentum conservation to ${\cal A}_{81}$. It is easy to show that the first term of ${\cal A'}_{1}$ of $<V_{C^{-1}} V_{\phi^{-1}} V_{\phi^{0}} V_{\phi^{0}}>$ can be reconstructed by the prescribed  ${\cal A}_{81}$.  Once more we  did apply   momentum conservation to ${\cal A}_{71}$ of $<V_{C^{-2}} V_{\phi^{0}} V_{\phi^{0}} V_{\phi^{0}}>$ to derive
   \beqa
  i p.\xi_2\Tr(P_{-}\fsC_{(n-1)}M_p \Gamma^{kbic}) \xi_{1i}\xi_{3k}Q_1 (-k_{3c}-k_{2c}-p_{c})k_{3b}\label{bb67}\eeqa
   where due to antisymmetric property of $\epsilon $ tensor the  first term in \reef{bb67} has no contribution to the contact interactions and the 3rd term does generate  the 2nd contact term interaction of ${\cal A'}_{1}$, meanwhile the 2nd term in  \reef{bb67}  will be an extra contact interaction in asymmetric picture for which we consider it  in the next section.

\vskip.1in

     Holding the same arguments  to ${\cal A}_{41}$ we get to obtain
   \beqa
  i p.\xi_3\Tr(P_{-}\fsC_{(n-1)}M_p \Gamma^{jaic}) \xi_{1i}\xi_{2j} Q_1 (-k_{2c}-k_{3c}-p_{c})k_{2a}\label{bb00}\eeqa
 needless to say that  the  first term in  \reef{bb00} has no contribution to amplitude and the 3rd term above can produce  precisely 3rd contact term interaction of ${\cal A'}_{1}$
    while the 2nd term  in above equation should be regarded as an extra contact interaction in asymmetric picture that will be taken into account in the next section.

\vskip.1in

     Finally let us produce the last contact interaction of  ${\cal A'}_{1}$
   of symmetric picture as follows. In order to do so, one has to apply momentum conservation to the 2nd term of ${\cal A}_{3}$ of asymmetric picture  to reach  to
    \beqa
  -i p.\xi_2 p.\xi_3Q_1 \xi_{1i}\Tr(P_{-}\fsC_{(n-1)}M_p \Gamma^{ic})  (-k_{2c}-k_{3c}-p_{c})\label{bb88}\eeqa

   where the last term in \reef{bb88} produces the the last contact interaction of  ${\cal A'}_{1}$
   of symmetric picture (the 4th term of  ${\cal A'}_{1}$) meanwhile the other terms in \reef{bb88} remain to be extra contact interactions in asymmetric amplitude.
\vskip.1in

Henceforth, up to now we have been able to precisely construct or generate  the entire contact terms that have been appearing in symmetric picture. However, as we have revealed  the other terms of asymmetric S-matrix elements lead us to conclude that those terms are extra contact interactions that can just be explored in asymmetric picture of the amplitude for which we are going to consider them in the next section.

  \subsection{ All order contact terms of Asymmetric S-Matrix  }

In previous section we compared the contact terms of the S-matrix element of an RR and three transverse scalars in both symmetric and asymmetric pictures. This leads us to explore various new contact interactions in an asymmetric picture of the amplitude. Therefore without further explanations we first write down all the other contact interactions that are just appeared in asymmetric picture as follows
\beqa
{\cal A}^{C^{-2}\phi^{0} \phi^{0} \phi^{0} }&=&{\cal A''}_{11}+{\cal A''}_{22}+{\cal A''}_{3}+{\cal A''}_{421}+{\cal A''}_{52}
+{\cal A''}_{5}\nonumber\\&&{\cal A''}_{61}+{\cal A''}_{621}+{\cal A''}_{71}+{\cal A''}_{721}
\labell{7114uu9}\eeqa
where
\beqa
{\cal A''}_{11}&\!\!\!\sim\!\!\!&i\Tr(P_{-}\fsC_{(n-1)}M_p)
\bigg[p.\xi_1 p.\xi_2 p.\xi_3Q_1\bigg],
\nonumber\\
{\cal A''}_{22}&\sim&i k_{2a}\xi_{2j}\Tr(P_{-}\fsC_{(n-1)}M_p \Gamma^{ja})
\bigg\{-p.\xi_1 p.\xi_3Q_1\bigg\}\nonumber\\
{\cal A''}_{3}&\sim&i (-k_{2c}-k_{3c})\xi_{1i}\Tr(P_{-}\fsC_{(n-1)}M_p \Gamma^{ic})
\bigg\{-p.\xi_2 p.\xi_3Q_1\bigg\}\nonumber\\
{\cal A''}_{5}&\sim&i\Tr(P_{-}\fsC_{(n-1)}M_p \Gamma^{jaic})\xi_{2j}\xi_{1i} p.\xi_3(-k_{3c})k_{2a}Q_1\nonumber\\
{\cal A''}_{421}&\sim& ip.\xi_3 L_2\bigg\{t\xi_{1i}\xi_{2j}\Tr(P_{-}\fsC_{(n-1)}M_p \Gamma^{ji})\bigg\}\nonumber\\
%-2\xi_{1}.\xi_{2}\Tr(P_{-}\fsC_{(n)}M_p \Gamma^{ac})(-k_{3c})k_{2a}\bigg\}\nonumber\\
{\cal A''}_{52}&\sim&i k_{3b}\xi_{3k}\Tr(P_{-}\fsC_{(n-1)}M_p \Gamma^{kb})
\bigg\{-p.\xi_2 p.\xi_1Q_1\bigg\}\nonumber\\
{\cal A''}_{61}&\sim&i\Tr(P_{-}\fsC_{(n-1)}M_p \Gamma^{kbja})\xi_{2j}\xi_{3k} p.\xi_1k_{3b}k_{2a}Q_1\nonumber\\
{\cal A''}_{621}&\sim& ip.\xi_1 L_5\bigg\{u\xi_{2j}\xi_{3k}\Tr(P_{-}\fsC_{(n-1)}M_p \Gamma^{kj})\bigg\}\nonumber\\
%-2\xi_{2}.\xi_{3}\Tr(P_{-}\fsC_{(n)}M_p \Gamma^{bc})k_{2a}k_{3b}\bigg\}\nonumber\\
{\cal A''}_{71}&\sim&i\Tr(P_{-}\fsC_{(n-1)}M_p \Gamma^{kbic})\xi_{1i}\xi_{3k} p.\xi_2k_{3b}(-k_{2c})Q_1\nonumber\\
{\cal A''}_{721}&\sim& ip.\xi_2 L_3\bigg\{-s\xi_{1i}\xi_{3k}\Tr(P_{-}\fsC_{(n-1)}M_p \Gamma^{ki})\bigg\}\nonumber\\
%+2\xi_{1}.\xi_{3}\Tr(P_{-}\fsC_{(n)}M_p \Gamma^{bc})(-k_{2c})k_{3b}\bigg\}
%\labell{4856}
\nonumber
\eeqa
%where the functions
% \beqa
%Q_1&=&(2)^{-2(t+s+u)+1}\pi{\frac{\Gamma(-u+\frac{1}{2})
%\Gamma(-s+\frac{1}{2})\Gamma(-t+\frac{1}{2})\Gamma(-t-s-u+1)}
%{\Gamma(-u-t+1)\Gamma(-t-s+1)\Gamma(-s-u+1)}},\nonumber\\
%L_2&=&(2)^{-2(t+s+u)}\pi{\frac{\Gamma(-u+1)
%\Gamma(-s+1)\Gamma(-t)\Gamma(-t-s-u+\frac{1}{2})}
%{\Gamma(-u-t+1)\Gamma(-t-s+1)\Gamma(-s-u+1)}}
%\nonumber\\
%L_3&=&(2)^{-2(t+s+u)}\pi{\frac{\Gamma(-u+1)
%\Gamma(-s)\Gamma(-t+1)\Gamma(-t-s-u+\frac{1}{2})}{\Gamma(-u-t+1)
%\Gamma(-t-s+1)\Gamma(-s-u+1)}}
%,\nonumber\\
%L_5&=&(2)^{-2(t+s+u)}\pi{\frac{\Gamma(-u)
%\Gamma(-s+1)\Gamma(-t+1)\Gamma(-t-s-u+\frac{1}{2})}
%{\Gamma(-u-t+1)\Gamma(-t-s+1)\Gamma(-s-u+1)}}
%\nonumber
%\eeqa

where these contact terms can be generated in a new standard form of effective field theory couplings that we address it now.
Note that all the leading term of the above couplings appear to be at $(\alpha')^3$ order. One can obtain their all order 
$\alpha'$ contact interactions as well.
\vskip.1in

 The expansion for $Q_1$ is given in Appendix 2.  If we  consider ${\cal A''}_{11}$, extract the trace and employ the Taylor expansion for all three scalar fields,  one can show that the leading contact interaction can be precisely obtained by the following coupling
\beqa
S_{14} &=& i(2\pi\alpha')^3\mu_p\int d^{p+1}\s
{1\over (p+1)!}(\veps^v)^{a_0\cdots a_{p}}
\bigg(\partial_i\partial_j\partial_k C^{(p+1)}_{ a_{0}... a_{p}}\phi^i\phi^j\phi^k \bigg)
\label{rrmm66}
\eeqa
 Now all ${\cal A''}_{22}$,${\cal A''}_{52}$ and ${\cal A''}_{3}$  have the same structure so we just produce ${\cal A''}_{22}$.  Therefore to produce this term in an EFT, one needs to  extract the trace and show that the second scalar comes from pull-back, while the first and third scalar come from the Taylor expansion.  Hence, the leading contact interaction can be precisely derived by the following coupling
\beqa
S_{15} &=& i(2\pi\alpha')^3\mu_p\int d^{p+1}\s
{1\over (p)!}(\veps^v)^{a_0\cdots a_{p}}
\bigg(D_{a_{0}} \phi^j \partial_i\partial_k C^{(p+1)}_{j a_{1}... a_{p}}\phi^i\phi^k \bigg)
\label{bbv3}
\eeqa

All order $\alpha'$ corrections to \reef{bbv3} can be derived by considering all the infinite terms appearing in the expansion of $Q_1$ as shown in Appendix 2.

\vskip.1in

Note that all ${\cal A''}_{5}$,${\cal A''}_{61}$ and ${\cal A''}_{71}$  have the same structure so we just produce ${\cal A''}_{5}$. Therefore for this term  one needs  to first  extract the trace and show that the first and the second scalar come from pull-back, while this time the  third scalar comes from the Taylor expansion.  Thus the leading contact interaction can be explored by the following coupling
\beqa
S_{16} &=& i(2\pi\alpha')^3\mu_p\int d^{p+1}\s
{1\over (p-1)!}(\veps^v)^{a_0\cdots a_{p}}
\bigg(D_{a_{0}} \phi^i D_{a_{1}} \phi^j\partial_k C^{(p+1)}_{ij a_{2}... a_{p}}\phi^k \bigg)
\nonumber
\eeqa
where its  all order $\alpha'$ corrections have already been given in Appendix 2.

Finally  all ${\cal A''}_{421}$,${\cal A''}_{621}$ and ${\cal A''}_{721}$  have the same structure and also $tL_2=sL_3=uL_5=L_{22}$,
 the expansion of $L_{22}$ is also given is Appendix 2. We just produce ${\cal A''}_{421}$. For this term  one needs  to first  extract the trace and show that the third scalar comes from Taylor expansion,  while the first and second scalar neither come from Taylor nor pull-back. The leading contact interaction can be precisely derived by the following coupling 
\beqa
S_{17} &=& i(2\pi\alpha')^3\mu_p\int d^{p+1}\s
{1\over (p+1)!}(\veps^v)^{a_0\cdots a_{p}}
\bigg( \phi^i \phi^j\partial_k C^{(p+3)}_{ij a_{0}... a_{p}}\phi^k \bigg)
\label{vie99}
\eeqa
where its  all order $\alpha'$ corrections can also be derived  as we did in \reef{2ee}.
\vskip.2in

Therefore in this section not only did we derive new couplings of RR in an EFT approach but also we have been able to find out their all order $\alpha'$
 higher derivative interactions on both  world volume and transverse directions.

%%%%%%%%%%%%%%%%%%%%%%%%%%%%%%%%
\section{Conclusion}
%%%%%%%%%%%%%%%%%%%%%%%%%%%%%%%%

In this paper we have analyzed in detail the four and five point functions of the string theory, including an RR in an asymmetric picture and either two or three real transverse scalar fields. We also compared the exact form of the S-matrix in asymmetric picture with its own result in symmetric picture.  We have obtained two different kinds of new bulk singularity structures as well as various new couplings in asymmetric picture that are absent in its symmetric picture.  We have also  generalized their all order $\alpha'$ higher derivative interactions as well.

 \vskip.1in

These two different kinds of  bulk singularity structures of string amplitude are $u,t,s$ (appeared in \reef{newsins1},\reef{newsins}) as well as $(t+s+u)$-channel bulk singularity structures \reef{amphigh87} that can just be explored in an asymmetric picture of the amplitude.  These bulk singularity structures carry momentum of RR in transverse directions  ( that is, $p^i, p^j, p^k$ terms). Note that these terms could have been derived if winding modes ($w^i, w^j, w^k$ terms) were allowed in the vertex operator of RR. However, in the vertex of RR  in ten dimensions of spacetime there are no winding modes in both symmetric and asymmetric picture of RR.  

Indeed these terms of the amplitude whose momenta of RR are carried in transverse direction cannot be obtained even by T-duality
transformation in flat ten dimensions of space-time. Hence the  presence of RR makes computations complicated as was explained in  \cite{Hatefi:2012ve,Park:2008sg}.
Hence we just explore these new bulk singularities as well as new couplings in asymmetric picture of the amplitude. Indeed we are also able to produce these two different  bulk singularity structures of string amplitude in field theory by taking into account various new couplings in effective field theory side as well.   We think that the importance of these results will be provided in future research topics, such as all order Myers effect and various other subjects in type II super string theory \cite{Hatefi:2015mio}.

\vskip.1in

We have also observed that at the level of EFT the super gravity background fields in DBI action must be some functions of super Yang-Mills. We have also shown that some particular Taylor expansion for the background fields should be taken into account as was noticed in Dielectric effect \cite{Myers:1999ps}. These results might be important both in constructing higher point functions of string theory amplitudes as well as discovering symmetries or mathematical results behind scattering amplitudes. We hope to  address these issues in near future.

\renewcommand{\theequation}{A.\arabic{equation}}
 \setcounter{equation}{0}
  \section*{Appendix 1 }
All $n_{i}'s $ for the asymmetric amplitude are given by
\beqa
n_1&=&\bigg(\eta^{ac}(\Gamma^{ji}C^{-1})_{\alpha\beta}
+\eta^{ij}(\Gamma^{ac}C^{-1})_{\alpha\beta}\bigg),\nonumber\\
n_2&=&\bigg(-\eta^{ac}\eta^{ij}(C^{-1})_{\alpha\beta}\bigg),\nonumber\\
n_3&=&\bigg(\eta^{ab}(\Gamma^{kj}C^{-1})_{\alpha\beta}
+\eta^{jk}(\Gamma^{ba}C^{-1})_{\alpha\beta}\bigg),\nonumber\\
n_4&=&\bigg(-\eta^{ab}\eta^{jk}(C^{-1})_{\alpha\beta}\bigg),\nonumber\\
n_5&=&\bigg(\eta^{bc}(\Gamma^{ki}C^{-1})_{\alpha\beta}
+\eta^{ik}(\Gamma^{bc}C^{-1})_{\alpha\beta}\bigg),\nonumber\\
n_6&=&\bigg(-\eta^{bc}\eta^{ik}(C^{-1})_{\alpha\beta}\bigg),\nonumber\\
\eeqa
Indeed the correlation function of three currents with  two spin fields, does carry so many terms. One can further simplify it and write it down in a very compact formula as below
\beqa
I_8^{kbjaic}&=&<:S_{\al}(x_4):S_{\be}(x_5)::\psi^c\psi^i(x_1):\psi^a\psi^j(x_2):\psi^b\psi^k(x_3)>\label{33312}\eeqa
with the following ingredients
\beqa
I_8^{kbjaic}&=&
\bigg\{(\Gamma^{kbjaic}C^{-1})_{{\alpha\beta}}+\alpha' n_7\frac{Re[x_{14}x_{25}]}{x_{12}x_{45}}+\alpha' n_8\frac{Re[x_{14}x_{35}]}{x_{13}x_{45}}+\alpha' n_9\frac{Re[x_{24}x_{35}]}{x_{23}x_{45}}
+\alpha'^2 n_{10}\nonumber\\&&\times
\bigg(\frac{Re[x_{14}x_{25}]}{x_{12}x_{45}}\bigg)^{2}
+\alpha'^2 n_{11}\bigg(\frac{Re[x_{14}x_{25}]}{x_{12}x_{45}}\bigg)\bigg(\frac{Re[x_{14}x_{35}]}{x_{13}x_{45}}\bigg)
+\alpha'^2 n_{12}\bigg(\frac{Re[x_{14}x_{25}]}{x_{12}x_{45}}\bigg)\bigg(\frac{Re[x_{24}x_{35}]}{x_{23}x_{45}}\bigg)\nonumber\\&&
+\alpha'^2 n_{13}\bigg(\frac{Re[x_{14}x_{35}]}{x_{13}x_{45}}\bigg)^{2}
+\alpha'^2 n_{14}\bigg(\frac{Re[x_{24}x_{35}]}{x_{23}x_{45}}\bigg)^{2}
+\alpha'^2 n_{15}\bigg(\frac{Re[x_{14}x_{35}]}{x_{13}x_{45}}\bigg)\bigg(\frac{Re[x_{24}x_{35}]}{x_{23}x_{45}}\bigg)\bigg\}
\label{hh333}\nonumber\\&&\times
2^{-3}x_{45}^{7/4}(x_{14}x_{15}x_{24}x_{25}x_{34}x_{35})^{-1}\nonumber
\eeqa
%%%
where
%%%
\beqa
%n_1&=&\bigg(\eta^{ac}(\Gamma^{ji}C^{-1})_{\alpha\beta}
%+\eta^{ij}(\Gamma^{ac}C^{-1})_{\alpha\beta}\bigg),\nonumber\\
%n_2&=&\bigg(-\eta^{ac}\eta^{ij}(C^{-1})_{\alpha\beta}\bigg),\nonumber\\
%n_3&=&\bigg(\eta^{ab}(\Gamma^{kj}C^{-1})_{\alpha\beta}
%+\eta^{jk}(\Gamma^{ba}C^{-1})_{\alpha\beta}\bigg),\nonumber\\
%n_4&=&\bigg(-\eta^{ab}\eta^{jk}(C^{-1})_{\alpha\beta}\bigg),\nonumber\\
%n_5&=&\bigg(\eta^{bc}(\Gamma^{ki}C^{-1})_{\alpha\beta}
%+\eta^{ik}(\Gamma^{bc}C^{-1})_{\alpha\beta}\bigg),\nonumber\\
%n_6&=&\bigg(-\eta^{bc}\eta^{ik}(C^{-1})_{\alpha\beta}\bigg),\nonumber\\
n_7&=&\bigg(\eta^{ac}(\Gamma^{kbji}C^{-1})_{\alpha\beta}
+\eta^{ij}(\Gamma^{kbac}C^{-1})_{\alpha\beta}\bigg),\nonumber\\
n_8&=&\bigg(\eta^{bc}(\Gamma^{kjai}C^{-1})_{\alpha\beta}
+\eta^{ik}(\Gamma^{bjac}C^{-1})_{\alpha\beta}\bigg),\nonumber\\
n_9&=&\bigg(\eta^{ab}(\Gamma^{kjic}C^{-1})_{\alpha\beta}
+\eta^{jk}(\Gamma^{baic}C^{-1})_{\alpha\beta}\bigg),\nonumber\\
n_{10}&=&\bigg(-\eta^{ac}\eta^{ij}(\Gamma^{kb}C^{-1})_{\alpha\beta}\bigg),\nonumber\\
n_{11}&=&\bigg(-\eta^{ac}\eta^{ik}(\Gamma^{bj}C^{-1})_{\alpha\beta}+\eta^{bc}\eta^{ij}(\Gamma^{ka}C^{-1})_{\alpha\beta}\bigg),\nonumber\\
n_{12}&=&\bigg(\eta^{ac}\eta^{jk}(\Gamma^{bi}C^{-1})_{\alpha\beta}-\eta^{ab}\eta^{ij}(\Gamma^{kc}C^{-1})_{\alpha\beta}\bigg),\nonumber\\
n_{13}&=&\bigg(-\eta^{bc}\eta^{ik}(\Gamma^{ja}C^{-1})_{\alpha\beta}\bigg),\nonumber\\
n_{14}&=&\bigg(-\eta^{ab}\eta^{jk}(\Gamma^{ic}C^{-1})_{\alpha\beta}\bigg),\nonumber\\
n_{15}&=&\bigg(-\eta^{bc}\eta^{jk}(\Gamma^{ai}C^{-1})_{\alpha\beta}+\eta^{ik}\eta^{ab}(\Gamma^{jc}C^{-1})_{\alpha\beta}\bigg)\nonumber
\eeqa

The integral could be carried over just in terms of  Gamma functions and no longer any hypergeometric function appears, where one needs to employ  the following sort of integrations on upper half plane \cite{Fotopoulos:2001pt}:
\beqa
 \int d^2 \!z |1-z|^{a} |z|^{b} (z - \bar{z})^{c}
(z + \bar{z})^{d}\nonumber
\eeqa
where   $a,b,c$ are arbitrary Mandelstam variable where for $ d= 0,1$ the result is given \cite{Fotopoulos:2001pt} and for $d=2$  one needs to work out \cite{Hatefi:2012wj} with the following Mandelstam definitions
\beqa
s&=&\frac{-\alpha'}{2}(k_1+k_3)^2,\quad t=\frac{-\alpha'}{2}(k_1+k_2)^2,\quad u=\frac{-\alpha'}{2}(k_2+k_3)^2
\nonumber\eeqa

 \section*{Appendix 2 }
The functions
 $Q_1,Q_2,Q_3,Q_4,Q_5,Q_6,Q_7,L_2,L_3,L_5,L_6$ are given by
\beqa
Q_1&=&(2)^{-2(t+s+u)+1}\pi{\frac{\Gamma(-u+\frac{1}{2})
\Gamma(-s+\frac{1}{2})\Gamma(-t+\frac{1}{2})\Gamma(-t-s-u+1)}
{\Gamma(-u-t+1)\Gamma(-t-s+1)\Gamma(-s-u+1)}},\nonumber\\
Q_2&=&(2)^{-2(t+s+u)-1}\pi{\frac{\Gamma(-u+\frac{1}{2})
\Gamma(-s+\frac{1}{2})\Gamma(-t-\frac{1}{2})\Gamma(-t-s-u)}
{\Gamma(-u-t)\Gamma(-t-s)\Gamma(-s-u+1)}},\nonumber\\
Q_3&=&(2)^{-2(t+s+u)-1}\pi{\frac{\Gamma(-u+\frac{1}{2})
\Gamma(-s-\frac{1}{2})\Gamma(-t+\frac{1}{2})\Gamma(-t-s-u)}
{\Gamma(-u-t+1)\Gamma(-t-s)\Gamma(-s-u)}},\nonumber\\
Q_4&=&(2)^{-2(t+s+u)-1}\pi{\frac{\Gamma(-u-\frac{1}{2})
\Gamma(-s+\frac{1}{2})\Gamma(-t+\frac{1}{2})\Gamma(-t-s-u)}
{\Gamma(-u-t)\Gamma(-t-s+1)\Gamma(-s-u)}},\nonumber\\
L_2&=&(2)^{-2(t+s+u)}\pi{\frac{\Gamma(-u+1)
\Gamma(-s+1)\Gamma(-t)\Gamma(-t-s-u+\frac{1}{2})}
{\Gamma(-u-t+1)\Gamma(-t-s+1)\Gamma(-s-u+1)}}
\nonumber\\
Q_5&=&(2)^{-2(t+s+u)-1}\pi{\frac{\Gamma(-u+\frac{1}{2})
\Gamma(-s+\frac{1}{2})\Gamma(-t-\frac{1}{2})\Gamma(-t-s-u)}
{\Gamma(-u-t+1)\Gamma(-t-s+1)\Gamma(-s-u+1)}}
\nonumber\\
\nonumber\\
Q_6&=&(2)^{-2(t+s+u)-1}\pi{\frac{\Gamma(-u-\frac{1}{2})
\Gamma(-s+\frac{1}{2})\Gamma(-t+\frac{1}{2})\Gamma(-t-s-u)}
{\Gamma(-u-t+1)\Gamma(-t-s+1)\Gamma(-s-u+1)}}
\nonumber\\
Q_7&=&(2)^{-2(t+s+u)-1}\pi{\frac{\Gamma(-u+\frac{1}{2})
\Gamma(-s-\frac{1}{2})\Gamma(-t+\frac{1}{2})\Gamma(-t-s-u)}
{\Gamma(-u-t+1)\Gamma(-t-s+1)\Gamma(-s-u+1)}}
\nonumber\\
%L_1&=&(2)^{-2(t+s+u)+1}\pi{\frac{\Gamma(-u+\frac{1}{2})
%\Gamma(-s+\frac{1}{2})\Gamma(-t+\frac{1}{2})\Gamma(-t-s-u+1)}
%{\Gamma(-u-t+1)\Gamma(-t-s+1)\Gamma(-s-u+1)}},\nonumber\\
L_3&=&(2)^{-2(t+s+u)}\pi{\frac{\Gamma(-u+1)
\Gamma(-s)\Gamma(-t+1)\Gamma(-t-s-u+\frac{1}{2})}{\Gamma(-u-t+1)
\Gamma(-t-s+1)\Gamma(-s-u+1)}}
,\nonumber\\
L_5&=&(2)^{-2(t+s+u)}\pi{\frac{\Gamma(-u)
\Gamma(-s+1)\Gamma(-t+1)\Gamma(-t-s-u+\frac{1}{2})}
{\Gamma(-u-t+1)\Gamma(-t-s+1)\Gamma(-s-u+1)}}
,\nonumber\\
L_6&=&(2)^{-2(t+s+u)-1}\pi{\frac{\Gamma(-u+\frac{1}{2})
\Gamma(-s+\frac{1}{2})\Gamma(-t+\frac{1}{2})\Gamma(-t-s-u)}
{\Gamma(-u-t+1)\Gamma(-t-s+1)\Gamma(-s-u+1)}},
\label{Ls}
\eeqa
 The expansion of $Q_1$ is 
\beqa
 Q_1 &=&-\frac{2\pi^{5/2}}{3}\left( 3\sum_{n=0}^{\infty}c_n(s+t+u)^{n+1}\right.
\left.+\sum_{n,m=0}^{\infty}c_{n,m}[(s^n t^m +s^m t^n)+(s^n u^m +s^m u^n)+(u^n t^m +u^m t^n)]\right.\nonumber\\
&&\left.+\sum_{p,n,m=0}^{\infty}f_{p,n,m}(s+t+u)^{p+1}[(s+t)^{n}(st)^{m}+(s+u)^{n}(su)^{m}+(u+t)^{n}(ut)^{m}]\right),\label{expansion11}
%L_2'&=&-\pi^{3/2}\sum_{n=-1}^{\infty}b_n\bigg(\frac{1}{3t}(u+s)^{n+1}+\frac{1}{3u}(t+s)^{n+1}
%+\frac{1}{3s}(u+t)^{n+1}\bigg)\nonumber\\&&+\sum_{p,n,m=0}^{\infty}e_{p,n,m}t^{p}(su)^{n}(s+u)^m\nonumber
%.\labell{high}
\eeqa

One can obtain  all order $\alpha'$ corrections to \reef{rrmm66} by considering all the higher derivative terms appearing in the expansion of $Q_1$ as explained in the following.
\beqa
(st)^{m}C\phi_1\phi_2\phi_3&=&(\alpha')^{2m}C D_{a_1}\cdots D_{a_{2m}}\phi_1 D^{a_{1}}\cdots D^{a_{m}}\phi_2
D^{a_{m+1}}\cdots D^{a_{2m}}\phi_3,\nonumber\\
(s+t)^{n}C\phi_1\phi_2\phi_3&=&(\alpha')^{n} C  D_{a_1}\cdots D_{a_{n}} \phi_1 D^{a_{1}}\cdots D^{a_{n}}(\phi_2
\phi_3),\nonumber\\
(s)^{m}t^n C\phi_1\phi_2\phi_3&=&(\alpha')^{n+m}C D_{a_1}\cdots D_{a_{n}} D_{a_{1}}\cdots D_{a_{m}}\phi_1 D^{a_{1}}\cdots D^{a_{n}}\phi_2 D^{a_{1}}\cdots D^{a_{m}}\phi_3,
\nonumber\\
(s)^{n}t^m  C\phi_1\phi_2\phi_3&=&(\alpha')^{n+m} C D_{a_1}\cdots D_{a_{n}} D_{a_{1}}\cdots D_{a_{m}}\phi_1 D^{a_{1}}\cdots D^{a_{m}}\phi_2 D^{a_{1}}\cdots D^{a_{n}}\phi_3,
\nonumber\\
(s+t+u)^{p+1} C \phi_1\phi_2\phi_3&=&(\frac{\alpha'}{2})^{p+1} C (D_{a}D^{a})^{p+1}(\phi_1\phi_2\phi_3).
\labell{2ee}
\eeqa
where the coefficients can be found in \cite{Hatefi:2012zh} and in the above equation all the commutator terms should not be considered. The expansion for $L_{22}$ is
\beqa
L_{22} &=&\frac{\pi^{3/2}}{3}\bigg(\sum_{n=-1}^{\infty}b_n (s+u)^{n+1} 
            +\sum_{p,n,m=0}^{\infty}e_{p,n,m}t^{p+1}(us)^{n}(u+s)^m\nonumber\\&&
            +\sum_{n=-1}^{\infty}b_n (s+t)^{n+1}
            +\sum_{p,n,m=0}^{\infty}e_{p,n,m}u^{p+1}(ts)^{n}(t+s)^m\nonumber\\&&
            +\sum_{n=-1}^{\infty}b_n (u+t)^{n+1}
            +\sum_{p,n,m=0}^{\infty}e_{p,n,m}s^{p+1}(tu)^{n}(t+u)^m\bigg)
 \label {expansion228}\eeqa
  %%%%%%%%%%%%%%%%%%%%%%%%%%%%%%%%%%%%%%%%%%%%%%%%%%%%%%%%%%%%%%%%%%%%%%%%%%%%%%%%%%%%%%%

\section*{Acknowledgments}

The author would like to thank  R. Russo, A. Sen, A. Brandhuber, W. Lerche,T.R. Taylor, C.Bachas, N.Arkani-Hamed,  K.Narain, L.Alvarez-Gaume, M.Douglas, J. Polchinski, E. Witten, M. B. Green , J. Schwarz, W. Siegel and  P.Vanhove for the correspondences and valuable comments/discussions. The initial stages of this work were performed in IAS in Princeton, at Simons Center , Stony brook, ICTP, IHES , UC Berkeley and  Ecole Normale Superieure at Paris, however,  most of the works have been  carried out at CERN during the author's visitorship. He would like to thank ICTP, Physics /Math departments of IHES.  Special thanks to theory division at CERN in Geneva for its hospitality. He is very grateful to W. Lerche , L.Alvarez-Gaume and CERN for their  help  throughout the completion of this work.

%%%%%%%%%%%%%%%%%%%%%%%%%%%%%%%%%%%%%%%%%%%%%%%%%%%%%%%%%%%%%%%%%%%%%%%%%%%%%%%%%%%%%%%

\end{document}